\documentclass[%
superscriptaddress,
 amsmath,amssymb,
 aps,
 prx,
 twocolumn,
]{revtex4-2}

\usepackage[english]{babel}
\usepackage{xcolor}

\usepackage[letterpaper,top=2cm,bottom=2cm,left=3cm,right=3cm,marginparwidth=1.75cm]{geometry}

\usepackage{amsmath}
\usepackage{graphicx}
\usepackage[colorlinks=true, allcolors=blue]{hyperref}
\usepackage{cleveref}
\usepackage{hyperref}
\usepackage{float}

\begin{document}

\title{Macroparticles with different weights relax to different temperatures in Particle-In-Cell simulations}

\author{Remi Lehe}
\affiliation{Lawrence Berkeley National Laboratory, Berkeley, California, USA}

\author{Arianna Formenti}
\affiliation{Lawrence Berkeley National Laboratory, Berkeley, California, USA}

\author{Justin R. Angus}
\affiliation{Lawrence Livermore National Laboratory, Livermore, California, USA}

\author{Jean-Luc Vay}
\affiliation{Lawrence Berkeley National Laboratory, Berkeley, California, USA}

\begin{abstract}
A distinctive feature of Particle-In-Cell (PIC) simulations is the use of macroparticles, each representing many physical particles. The associated macroparticle weight can vary from one species to another, or even from one macroparticle to another. Here we show that, when macroparticles have different weights, the plasma evolves toward an unphysical thermal equilibrium in which the temperatures are unequal: lower-weight macroparticles reach a higher temperature than physically expected, and higher-weight macroparticles a lower one. We show that this unphysical equilibrium is difficult to avoid, but that reaching it takes time, and that the associated timescale depends on the effective collisionality of the PIC algorithm. We derive an equation that predicts the full time evolution of the temperatures, in good agreement with PIC simulations, and use it to identify strategies to delay the establishment of this unphysical thermal equilibrium.
\end{abstract}

\maketitle

\section{Introduction}

The Particle-In-Cell (PIC) algorithm \cite{Birdsall,Hockney} is a widely-used approach for simulating plasmas, with applications across diverse fields such as astrophysical systems, plasma confinement devices, plasma thrusters, and laser-plasma interactions. One distinctive feature of the PIC algorithm is the use of \emph{macroparticles}, each representing multiple physical particles. The number of physical particles represented by a given macroparticle is generally referred to as the \emph{macroparticle weight}.

Oftentimes, all the macroparticles in a given simulation have equal weight. However, there are also many cases where the weights differ, either from one particle species to another, or even from one macroparticle to another within a given species. Examples include:

\begin{itemize}
\item Simulations with a spatially-varying initial density. In this case, macroparticles are sometimes initialized with a weight proportional to the initial density, and with a fixed number of macroparticles per cell, so as to represent the high-density regions with a manageable number of macroparticles.

\item Simulations using a cylindrical or spherical grid, in which case macroparticles are often initialized with a radius-dependent weight, so as to represent high-radius plasma with fewer macroparticles.

\item Plasmas that incorporate trace elements or impurities, in which case the macroparticles of the impurity species will often have a lower weight than the macroparticles of the background plasma. 

\item Simulations with low-probability Monte Carlo reactions (e.g., low-probability fusion reactions). In this case, the products of the reaction are often given lower weight compared to the reactants (and, accordingly, the probability of the reaction is increased by a corresponding \emph{production multiplier} \cite{Higginson2019Fusion,LavellPoP2024,LavellFrontiers2024}), so as to maintain a statistically-significant number of discrete reaction events. This includes for instance inertial fusion scenarios, where the alpha particles produced by the fusion reaction have lower weight than the background plasma.
 
\item Simulations that use macroparticle splitting or merging algorithms (e.g., \cite{Vranic,Smets}). Merging macroparticles into a smaller number of macroparticles with higher weights can be useful to control computational load, especially in simulations that generate new particles at an exponential rate \cite{Vranic}. On the other hand, splitting macroparticles into lower-weights can reduce stochastic noise in sparsely-populated regions of phase space (for instance, the tail of a Maxwellian distribution). In both cases, the weights generally differ from one macroparticle to another, and the distribution of weights evolves dynamically during the simulation.

\end{itemize}
In these instances, it is often assumed that the differing weights only affect the amount of random noise in the simulation, and that the average statistical properties of the plasma species (e.g., their temperatures) are largely independent of the ratio of macroparticle weights.

However, we show here that this is generally not the case. More specifically, we point out that there is a systematic, unphysical energy transfer from high-weight particles to low-weight macroparticles. As a result, at equilibrium, low-weight macroparticles will reach a higher temperature than physically expected, while high-weight macroparticles will reach a lower one.

As we will show, this unphysical energy transfer arises because the PIC algorithm naturally captures some of the physics of Coulomb collisions between charged macroparticles, albeit in a form distorted by the grid, macroparticle shape, and macroparticle weight. The existence of this effective, distorted collisionality was recognized as early as the 1970s \cite{Langdon,Okuda,Hockney1971} and is explained in reference textbooks on the PIC method \cite{Birdsall,Hockney}. Additionally, it has been noted that this effective collisionality can result in unphysical thermalization rates in PIC simulations \cite{Langdon,Melzani,Jubin,Park}. However, to our knowledge, the consequences of this effective collisionality for macroparticles of unequal weights, and in particular for their thermal equilibrium, have not been discussed previously.

The remainder of this paper is structured as follows. We begin by discussing the empirical observation of this effect in PIC simulations (\cref{sec:collisionlessempirical}), and then derive an equation that predicts both the thermal equilibrium state and the timescale over which it is reached (\cref{sec:collisionlesstheory}). This timescale depends on an effective Coulomb logarithm for PIC, which is further discussed in \cref{sec:approximate}.

\section{Observation of an unphysical, weight-dependent temperature equilibrium}
\label{sec:collisionlessempirical}

To illustrate the effect mentioned in the introduction, we perform 3D PIC simulations of a uniform plasma in a periodic box, using the code WarpX \cite{WarpX} and with parameters given in \cref{tab:parameters}. The plasma consists of electrons (labeled as \emph{species 1}) and ions (\emph{species 2}) initialized with equal density ($n_1 = n_2$) but different initial temperatures ($T_{1,\mathrm{init.}} = 80 \,\mathrm{eV}$, $T_{2,\mathrm{init.}} = 120\,\mathrm{eV}$). As is commonly done in numerical studies, the ions are given a low mass $m_2 = 10\,m_e$ for faster simulations; the impact of the ion mass on the relevant timescales is further discussed in \cref{sec:approximate}. We run different cases, whereby the number of macroparticles per cell (and thus the macroparticle weight) is varied independently for species 1 and 2, while preserving the same physical density ($n_1 = n_2 = 4 \times 10^{26}$ m$^{-3}$) as summarized in \cref{tab:parameters3D}. 

\begin{figure*}
\includegraphics[width=\linewidth]{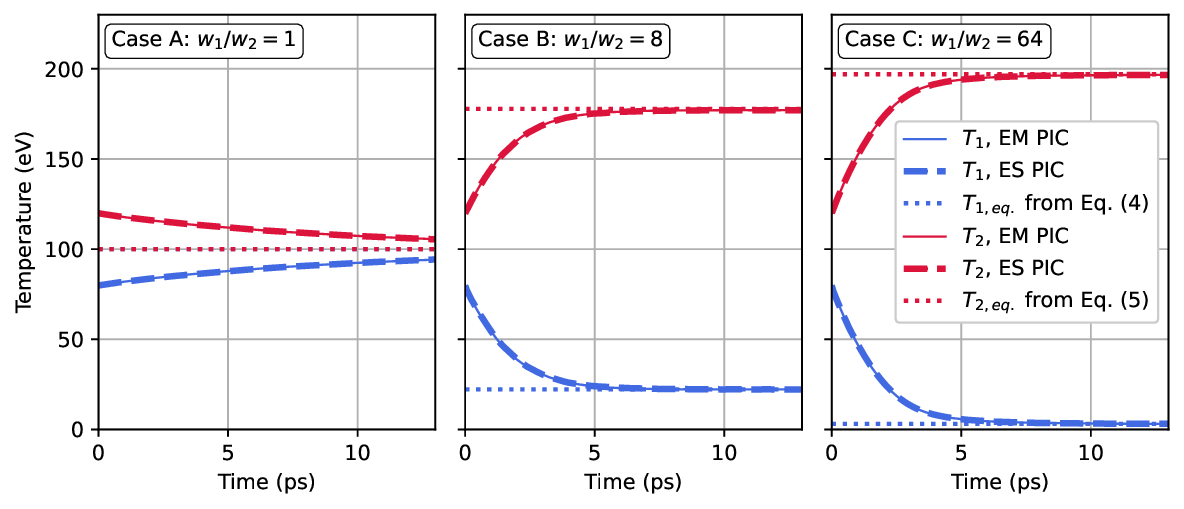}
\caption{Evolution of the temperature of each species ($T_1$ in blue and $T_2$ in red) in PIC simulations of a uniform plasma with parameters given in \cref{tab:parameters}, and with different ratios of macroparticle weights $w_1/w_2$: see case A/B/C in \cref{tab:parameters3D}. The lines represent the results of electromagnetic PIC simulations (EM PIC, solid line) and electrostatic PIC simulations (ES PIC, dashed line) respectively, for which the temperatures were extracted using \cref{eq:extract_T}. The dotted lines represent the predictions of \cref{eq:twospeciesequilibrium1,eq:twospeciesequilibrium2}.}
\label{fig:weights}
\end{figure*}

\begin{table}
  \begin{tabular}{c|c}
    \hline
    Parameter & Value\\
    \hline
    Physical density $n_1 = n_2$ & $4 \times 10^{26}$ m$^{-3}$  \\
    Properties of species 1 & $m_1$ = $m_e$, $q_1 = -e$ \\
    Properties of species 2 & $m_2$ = $10\,m_e$, $q_2 = +e$ \\
    Timestep $\Delta t$ & 0.034 $\omega_{p,e}^{-1}$ = 0.03 fs \\
    Cell size $\Delta x = \Delta y = \Delta z$ & 0.059 $c/\omega_{p,e}$ = 15.6 nm\\
    Number of cells (3D grid) & $16\times16\times16$ \\
    \hline
  \end{tabular}
  \caption{Physical and numerical parameters of the simulations discussed in \cref{sec:collisionlessempirical}}
  \label{tab:parameters}
\end{table}

\begin{table*}
\begin{center}
  \begin{tabular}{c|c|c}
    \hline
    Case & Species 1 (electrons) & Species 2 (ions) \\
    \hline
    Case A & 64 macroparticles/cell ($w_1 = 23.8$) & 64 macroparticles/cell ($w_2 = 23.8$)  \\
    Case B & 8 macroparticles/cell ($w_1 = 191$) & 64 macroparticles per cell ($w_2 = 23.8$)  \\
    Case C & 8 macroparticles/cell ($w_1 = 191$) & 512 macroparticles/cell ($w_2 = 2.98$)\\
    \hline
  \end{tabular}  
\end{center}  
  \caption{Number of macroparticles per cell (and corresponding weights) for the simulations displayed in \cref{fig:weights}
  \label{tab:parameters3D}
  }
\end{table*}

For each case in \cref{tab:parameters3D}, we perform two types of simulations: one using the electromagnetic (EM) PIC algorithm and the other using the electrostatic (ES) PIC algorithm. Both simulations use standard explicit PIC methods. The EM PIC simulation uses the Finite-Difference Time-Domain (FDTD) Yee solver to update the fields \cite{Yee}, and the Esirkepov charge-conserving algorithm for current deposition \cite{Esirkepov}. The ES PIC simulation uses standard second-order finite-difference discretization of the Poisson equation $\boldsymbol{\nabla}^2 \phi = -\rho/\epsilon_0$, which is solved on the nodes of the grid at each timestep using the multigrid method \cite{multigrid}. The electrostatic field $\boldsymbol{E} = -\boldsymbol{\nabla}\phi$ is obtained from $\phi$ on the staggered Yee grid with centered, second-order finite-difference. Both the EM and ES PIC simulations use the Boris pusher \cite{Boris} and no smoothing of the deposited charge or current. Importantly, all simulations use the ``energy-conserving'' gathering scheme \cite{Birdsall,Lewis} and splines of order 3 for the macroparticle shape factor. Finally, the simulations do \emph{not} use any Monte Carlo collision module \cite{TA77,Nanbu1997,Nanbu1998}, and thus the only possible source of collisionality is the self-consistent interaction between charged particles captured by the PIC algorithm itself.

\Cref{fig:weights} shows the evolution of the temperature of each species, computed using the definition of the temperature as the second-order moment of $\boldsymbol{v}$ (assuming a plasma with zero drift, i.e. $\langle \boldsymbol{v} \rangle_\sigma = 0$): 
\begin{equation}
k_B T_{\sigma} \equiv \frac{m_{\sigma}}{3}\langle \boldsymbol{v}^2\rangle_\sigma \label{eq:def_temperature}
\end{equation}
where $\sigma = 1,2$ is the species index, $m_\sigma$ is the mass of the \emph{physical} particles of species $\sigma$, and $\langle ... \rangle_\sigma$ denotes an average over the population of species $\sigma$. More precisely, this average is evaluated as a weighted sum over the $N_\sigma$ macroparticles of species $\sigma$: 
\begin{align}
k_B T_{\sigma} &= \frac{m_{\sigma}}{3}\frac{\sum_{p=1}^{N_\sigma} w_\sigma \boldsymbol{v}^2_p}{\sum_{p=1}^{N_\sigma}w_\sigma}  \\
&= \frac{m_{\sigma}}{3}\frac{\sum_{p=1}^{N_\sigma} \boldsymbol{v}^2_p}{N_\sigma} \label{eq:extract_T}
\end{align}
where the simplification in the second line follows from the fact that, here, all macroparticles \emph{within} a given species share the same weight $w_\sigma$. %This is indeed the standard way one would extract the physical temperature from a PIC simulation, and the resulting value is not expected to depend on the macroparticle weight.

If the plasma were purely collisionless, there would be no energy transfer between the two species, and each species would retain its initial temperature. However, because the PIC algorithm captures some of the physics of Coulomb collisions between charged macroparticles, the species are expected to exchange energy, causing their temperatures to evolve. Under these conditions, the expected physical behavior is of course that the system reaches an equilibrium in which the temperatures of species 1 and 2 are equal ($T_{1,\mathrm{eq.}} = T_{2,\mathrm{eq.}}$), regardless of the macroparticle weights.

Yet, \cref{fig:weights} reveals a striking result: although the system does evolve toward an equilibrium, the equilibrium temperatures of species 1 and 2 are not equal ($T_{1,\mathrm{eq.}} \neq T_{2,\mathrm{eq.}}$) whenever the macroparticle weights differ ($w_1 \neq w_2$). In addition, these equilibrium temperatures follow a simple, predictable relation, which is independent of the species masses $m_1$, $m_2$ and depends only on the ratio of the weights $w_1/w_2$:
\begin{align} 
T_{1,\mathrm{eq.}} &= \frac{1}{1 + w_1/w_2} (T_{1,\mathrm{init.}} + T_{2,\mathrm{init.}}) \label{eq:twospeciesequilibrium1} \\
T_{2,\mathrm{eq.}} &= \frac{1}{1 + w_2/w_1} (T_{1,\mathrm{init.}} + T_{2,\mathrm{init.}}) 
\label{eq:twospeciesequilibrium2}
\end{align}
The justification for this relation will become clear in \cref{sec:collisionlesstheory}.

This unphysical behavior is robustly observed across a range of conditions. As shown in \cref{fig:weights}, it appears in both electromagnetic and electrostatic PIC. In fact, the time evolution of the temperatures is identical in these two cases, suggesting that it is driven primarily by effects that are electrostatic in nature. Furthermore, although the simulations presented here are 3D, a similar unphysical equilibrium is also observed in 1D and 2D PIC simulations, as shown in \cref{app:2d1d}.

%A notable difference in these lower-dimensionality simulations is that only the velocity distribution along the axes of the simulation (e.g., in-plane velocities in 2D) evolves and relaxes towards the unphysical equilibrium \cref{eq:twospeciesequilibrium1,eq:twospeciesequilibrium2}, while the velocity distribution in the invariant directions (e.g., out-of-plane direction in 2D) decouples and remains unchanged.

Note that this unphysical effect is clearly distinct from the numerical heating or cooling that is sometimes observed in explicit PIC simulations. Indeed, although energy is exchanged between the two species here, the total energy in the simulation was found to be conserved to a high degree of accuracy. (The relative change in total energy, i.e. in the sum of the field and kinetic energy, remained below $10^{-5}$ for all simulations shown in \cref{fig:weights}.)

\section{Theoretical explanation}
\label{sec:collisionlesstheory}

It turns out that it is possible to predict this unphysical evolution of temperature theoretically, in the case where this evolution is primarily driven by electrostatic interactions. In this section, we first recall known results for the evolution of the distribution function in PIC codes (\cref{sec:evolution-distribution}). We then use these results to derive an equation for the evolution of the temperature (\cref{sec:evolution-temperature}) and compare its predictions with simulation results (\cref{sec:comparison}).

\subsection{Evolution of the distribution function $f_\sigma$}
\label{sec:evolution-distribution}

We consider a uniform plasma with electrostatic interactions between macroparticles. The plasma consists of several species (indexed by $\sigma$), each characterized by:
\begin{itemize}
\item the weight $w_\sigma$ of its macroparticles, which is assumed to be uniform \emph{within} one species, but may vary from one species to another;
\item the density, charge and mass of the \emph{physical} particles it represents, denoted $n_\sigma$, $q_\sigma$ and $m_\sigma$. With this notation, the density, charge and mass of the corresponding macroparticles are $n_{\mathrm{macro},\sigma} = n_\sigma/w_\sigma$, $q_{\mathrm{macro},\sigma} = w_\sigma q_\sigma$, $m_{\mathrm{macro},\sigma} = w_\sigma m_\sigma$;
\item its distribution function $f_\sigma(\boldsymbol{x}, \boldsymbol{v}, t)$, normalized such that $\int f_\sigma(\boldsymbol{x}, \boldsymbol{v}, t)\,d\boldsymbol{v} = 1$
\end{itemize}
More precisely, the distribution function is defined here so that $(n_\sigma/w_\sigma)\times f_\sigma(\boldsymbol{x}, \boldsymbol{v}, t)\,d\boldsymbol{x}\,d\boldsymbol{v}$ is the average number of macroparticles found in a phase space volume $d\boldsymbol{x}\,d\boldsymbol{v}$ centered at $\boldsymbol{x}, \boldsymbol{v}$. (This average is taken over a large number of \emph{independent realizations} of a PIC simulation, in which the initial position and velocity of each macroparticle are randomly drawn from $f_\sigma(\boldsymbol{x}, \boldsymbol{v}, t=0)$ for each independent realization \cite{Touati}.) Note that, because the plasma is assumed to be uniform, the distribution function is in fact independent of $\boldsymbol{x}$ and can be written $f_\sigma(\boldsymbol{v}, t)$.

For the ``energy-conserving'' scheme of electrostatic PIC (also referred to as Lewis' model in \cite{Birdsall}), the evolution of the distribution function $f_\sigma$ due to the effective PIC collisionality is given by \cref{eq:BGLequation,eq:BGLoperator1,eq:BGLoperator2,eq:dielectric} below. We note that these equations were historically given for a single species in \cite{Birdsall} (chapter 12, equation 19), but that they can be generalized to multiple species, as done for instance in \cite{Touati}.

\begin{widetext}
\begin{align}
\frac{\partial\, f_\sigma(\boldsymbol{v},t)}{\partial t} &= - \sum_{{\sigma'}}\frac{\partial \,}{\partial \boldsymbol{v}}\cdot \int d\boldsymbol{v}'\; \left[\frac{m_{\sigma'}}{m_\sigma} w_{\sigma'} \boldsymbol{Q}_{\sigma {\sigma'}}\cdot\frac{\partial \;}{\partial \boldsymbol{v}} - w_\sigma \boldsymbol{Q'}_{\sigma {\sigma'}}\cdot\frac{\partial \;}{\partial \boldsymbol{v'}} \right] f_\sigma(\boldsymbol{v},t)f_{\sigma'}(\boldsymbol{v'},t)
\label{eq:BGLequation}\\
\boldsymbol{Q}_{\sigma {\sigma'}} &= - \pi \frac{\omega_{p,\sigma}^2 \omega_{p,{\sigma'}}^2}{n_\sigma}\int_g\frac{d\boldsymbol{k}}{(2\pi)^3}\sum_{\boldsymbol{p} \boldsymbol{p'}}\frac{S_m^2(\boldsymbol{k}_{\boldsymbol{p}})S_m^2(\boldsymbol{k}_{\boldsymbol{p'}})}{|\epsilon(\boldsymbol{k}, \boldsymbol{k}_{\boldsymbol{p}}\cdot\boldsymbol{v})|^2}\frac{\boldsymbol{k}_{\boldsymbol{p}} \,\boldsymbol{k}_{\boldsymbol{p}}}{K^4(\boldsymbol{k})}
\delta(\boldsymbol{k}_{\boldsymbol{p}}\cdot\boldsymbol{v}-\boldsymbol{k}_{\boldsymbol{p'}}\cdot\boldsymbol{v'}, 2\pi/\Delta t ) \label{eq:BGLoperator1}
\\
\boldsymbol{Q'}_{\sigma {\sigma'}} &= - \pi \frac{\omega_{p,\sigma}^2 \omega_{p,{\sigma'}}^2}{n_\sigma}\int_g\frac{d\boldsymbol{k}}{(2\pi)^3}\sum_{\boldsymbol{p} \boldsymbol{p'}}\frac{S_m^2(\boldsymbol{k}_{\boldsymbol{p}})S_m^2(\boldsymbol{k}_{\boldsymbol{p'}})}{|\epsilon(\boldsymbol{k}, \boldsymbol{k}_{\boldsymbol{p}}\cdot\boldsymbol{v})|^2}\frac{\boldsymbol{k}_{\boldsymbol{p}}\,\boldsymbol{k}_{\boldsymbol{p'}}}{K^4(\boldsymbol{k})}
\delta(\boldsymbol{k}_{\boldsymbol{p}}\cdot\boldsymbol{v}-\boldsymbol{k}_{\boldsymbol{p'}}\cdot\boldsymbol{v'}, 2\pi/\Delta t )
\label{eq:BGLoperator2}
\\
\epsilon( \boldsymbol{k}, \omega) &= 1 + \sum_{\sigma''} \frac{\omega_{p,\sigma''}^2}{K^2(\boldsymbol{k})}\sum_{\boldsymbol{p''}} S_m^2(\boldsymbol{k}_{\boldsymbol{p''}})\int d\boldsymbol{v}'' \;\boldsymbol{k}_{\boldsymbol{p''}}\cdot\frac{\partial f_{\sigma''}(\boldsymbol{v}'',t)}{\partial \boldsymbol{v}''}\frac{\Delta t}{2}\cot\left[ (\omega - \boldsymbol{k}_{\boldsymbol{p''}}\cdot \boldsymbol{v}'')\frac{\Delta t}{2}\right]
\label{eq:dielectric}
\end{align}
\end{widetext}
where the sum over ${\sigma'}$ in \cref{eq:BGLequation} represents a sum over all species (including species $\sigma$ itself) and where $\boldsymbol{Q}_{\sigma {\sigma'}}, \boldsymbol{Q'}_{\sigma {\sigma'}}$ are similar to the physical Balescu-Guernsey-Lenard (BGL) collision operator \cite{Ichimaru}, but incorporate the unphysical effects of the discrete grid, discrete timestep and macroparticle shape factor. ($\boldsymbol{Q}_{\sigma {\sigma'}}$ and $\boldsymbol{Q'}_{\sigma {\sigma'}}$ differ only in one term $\boldsymbol{k}_{\boldsymbol{p}}$ replaced by $\boldsymbol{k}_{\boldsymbol{p'}}$.) In the above expressions, $\omega_{p,\sigma}^2 = n_\sigma q_\sigma^2/(m_\sigma \epsilon_0)$ is the plasma frequency associated with species $\sigma$. The integral sign $\int_g d\boldsymbol{k}$ represents integration over the first Brillouin zone associated with the simulation grid, and $\sum_{\boldsymbol{p}\boldsymbol{p'}}$ represents a sum over spatial aliases. $\boldsymbol{k}_{\boldsymbol{p}} = (k_x + 2\pi p_x/\Delta x, k_y + 2\pi p_y/\Delta y, k_z + 2\pi p_z/\Delta z)$ is a grid alias of vector $\boldsymbol{k}$. $S_m(\boldsymbol{k})$ is the Fourier transform of the macroparticle shape, when using splines of order $m$: $S_m(\boldsymbol{k}) = [\,\mathrm{sinc}(k_x\Delta x/2)\,\mathrm{sinc}(k_y\Delta y/2)\,\mathrm{sinc}(k_z\Delta z/2)\,]^{m+1}$, where $\mathrm{sinc}(x) \equiv \sin(x)/x$. In addition, $K^2(\boldsymbol{k})$ is the Fourier transform of the discrete Laplacian operator used in the Poisson equation; for the second-order stencil used here $K^2(\boldsymbol{k}) = k_x^2 \mathrm{sinc}(k_x\Delta x/2)^2 + k_y^2 \mathrm{sinc}(k_y\Delta y/2)^2 + k_z^2 \mathrm{sinc}(k_z\Delta z/2)^2$. By definition, the notation $\delta(\omega, 2\pi/\Delta t)$ is a sum over time aliases: $\delta(\omega, 2\pi/\Delta t) \equiv \sum_{q=-\infty}^{\infty}\delta(\omega - q 2\pi/\Delta t)$. Finally, $\epsilon(\boldsymbol{k}, \omega)$ is the dielectric function of the plasma. For more details on these quantities and notations, see section 9-5 in \cite{Birdsall}.

By comparing \cref{eq:BGLequation,eq:BGLoperator1,eq:BGLoperator2,eq:dielectric} with the physical BGL equation \cite{Ichimaru}, one can see that the PIC algorithm naturally exhibits some form of collisionality, but that it differs quantitatively from the physical collisionality predicted by the BGL equation
\begin{itemize}
\item The effect of collisions with impact parameter below the effective macroparticle size (which depends on the macroparticle shape) is suppressed; this tends to reduce the effective collision frequency. In \cref{eq:BGLoperator1,eq:BGLoperator2}, this is captured by the fact that $S_m(\boldsymbol{k})$ goes to zero for large $\boldsymbol{k}$, which cuts off the high $\boldsymbol{k}$ contributions to $\boldsymbol{Q}_{\sigma \sigma'}$ and $\boldsymbol{Q}'_{\sigma \sigma'}$.
\item On the other hand, the collisionality is artificially enhanced by the macroparticle weight. This stems from the fact that the collision frequency between charged particles with density $n$, charge $q$ and mass $m$ is proportional to $n q^4 /m^2$ \cite{Spitzer}. When applied to macroparticles, this quantity becomes $n_{\mathrm{macro},\sigma} q^4_{\mathrm{macro},\sigma}/m_{\mathrm{macro},\sigma}^2 = w_\sigma n_\sigma q_\sigma^4/m_\sigma^2$ and thus the collision frequency is enhanced by an unphysical factor $w_\sigma$ relative to that of the underlying physical particles. In \cref{eq:BGLequation}, this is captured by the terms $w_{\sigma}$ and $w_{\sigma'}$.
\end{itemize}

\subsection{Evolution of the temperature $T_\sigma$, and predicted equilibrium}
\label{sec:evolution-temperature}

Since \cref{eq:BGLequation,eq:BGLoperator1,eq:BGLoperator2,eq:dielectric} govern the evolution of the distribution function $f_{\sigma}(\boldsymbol{v}, t)$, the evolution of the temperature $T_\sigma(t)$ can be inferred by expressing the definition \cref{eq:def_temperature} of $T_\sigma$ as an average over the distribution function $f_{\sigma}(\boldsymbol{v}, t)$:
\begin{align}
k_B T_\sigma(t) &=
\frac{m_\sigma}{3} \int d\boldsymbol{v}\; f_\sigma(\boldsymbol{v}, t)\,\boldsymbol{v}^2\\
k_B\frac{d\, T_\sigma(t)}{dt} &= \frac{m_\sigma}{3} \int d\boldsymbol{v}\;  \frac{\partial  \,f_\sigma(\boldsymbol{v}, t)}{\partial t}\,\boldsymbol{v}^2
\label{eq:firstTequation}
\end{align}
The time derivative $\partial_t f_\sigma(\boldsymbol{v}, t)$ can then be replaced by the expression given in the BGL-like equation \cref{eq:BGLequation}. This procedure is carried out in \cref{app:derivationT}, with two additional simplifying assumptions:
\begin{itemize}
\item The distribution function of each species $f_{\sigma}$ is assumed to remain Maxwellian at all times, with a time-dependent temperature $T_{\sigma}(t)$:
\begin{equation}
f_{\sigma}(\boldsymbol{v}, t) = \left(\frac{m_{\sigma}}{2\pi k_B T_{\sigma}(t)}\right)^{3/2}\!\!\!\!\exp\left( -\frac{m_{\sigma}\boldsymbol{v}^2}{2k_B T_{\sigma}(t)}\right)
\label{eq:Maxwellian}
\end{equation}
\item The influence of temporal and spatial aliases in \cref{eq:BGLequation,eq:BGLoperator1,eq:BGLoperator2,eq:dielectric} is neglected.
\end{itemize}
Under these assumptions, the combination of \cref{eq:firstTequation} and \cref{eq:BGLequation,eq:BGLoperator1,eq:BGLoperator2,eq:dielectric} reduces to a relatively compact equation:
\begin{widetext}
\begin{equation}
\frac{d\, T_\sigma}{dt} =\sum_{{\sigma'}}  \frac{2\;\omega_{p,\sigma}^2 \omega_{p,{\sigma'}}^2}{3(2\pi)^{3/2} n_\sigma} \frac{ w_{\sigma'} T_{\sigma'} - w_\sigma T_\sigma }{\left[ \, k_B T_\sigma/m_\sigma + k_B T_{\sigma'}/m_{\sigma'}\,\right]^{3/2}}\log \Lambda_{PIC, \sigma \sigma'}
\label{eq:evolutionT}
\end{equation}
\end{widetext}

This result is very similar to the equation for temperature evolution in Spitzer theory \cite{Spitzer}, but with the additional influence of the weights in the numerator $(w_{\sigma'} T_{\sigma'} - w_\sigma T_\sigma)$, and with a modified Coulomb logarithm $\log \Lambda_{PIC, \sigma \sigma'}$ that represents the effective collisionality of the PIC algorithm, and takes into account the effect of the macroparticle shape and discrete grid.

This Coulomb logarithm has a somewhat complicated expression, which is given in \cref{eq:logPIC} below (see \cref{app:derivationT} for its derivation). In practice, this expression can be evaluated numerically (see \cref{sec:comparison}) or approximated for certain regimes (see \cref{sec:approximate}).
\begin{widetext}
\begin{equation}
\log \Lambda_{PIC, \sigma \sigma'} = \int_g\frac{d\boldsymbol{k}}{2\pi^{3/2}} \frac{S_m^4(\boldsymbol{k})k}{K^4(\boldsymbol{k})}\int_{-\infty}^\infty \!\!\!\!\!d\alpha\frac{\alpha^2e^{-\alpha^2}}{\left|\epsilon\left(\boldsymbol{k}, \frac{\alpha k}{\sqrt{ \frac{m_\sigma}{2k_B T_{\sigma}} + \frac{m_{\sigma'}}{2k_B T_{\sigma'}} }} \right) \right|^2}
\label{eq:logPIC}
\end{equation}
\begin{equation} 
\epsilon(\boldsymbol{k}, \omega) = 1 - \sum_{\sigma''} \frac{S_m^2(\boldsymbol{k})}{2K^2(\boldsymbol{k}) \lambda_{D,\sigma''}^2}  Z'\left( \frac{\omega}{k} \sqrt{\frac{m_{\sigma''}}{2 k_B T_{\sigma''}}}\,  \right)
\label{eq:dielectricM} 
\end{equation}
\end{widetext}
In \cref{eq:logPIC,eq:dielectricM}, we used again the notation $K(\boldsymbol{k})$ and $S_m(\boldsymbol{k})$ introduced in \cref{sec:evolution-distribution}. Additionally, $\lambda_{D,\sigma''} = \sqrt{k_B T_{\sigma''}/m_{\sigma''}\omega_{p,\sigma''}^2}$ is the Debye length associated with species $\sigma''$. In the expression of $\epsilon(\boldsymbol{k}, \omega)$, $Z'$ is the derivative of the plasma dispersion function \cite{FriedConte1961} (see also \cref{eq:plasma_dispersion}). While \cref{eq:evolutionT,eq:logPIC,eq:dielectricM} are written specifically for the case of 3D simulations, more generic expressions valid for 1D, 2D and 3D simulations are given in  \cref{eq:evolutionT_anyD,eq:logPIC_anyD,eq:dielectricM_anyD} of the appendix.

As will be shown in the next section, the predictions of \cref{eq:evolutionT,eq:logPIC,eq:dielectricM} are in very good agreement with PIC simulation results. Before turning to this comparison, it is worth first pointing out that, despite their complexity, \cref{eq:evolutionT,eq:logPIC,eq:dielectricM}, exhibit a number of simple properties.

First, multiplying \cref{eq:evolutionT} by a constant factor $n_\sigma\,3k_B/2$, summing over $\sigma$ and noticing that the right-hand side is anti-symmetric with respect to $\sigma$ and $\sigma'$ yields:
\begin{equation} 
\frac{d\,}{dt}\left(\sum_{\sigma} n_\sigma \frac{3 k_B T_\sigma}{2} \right)=0
\label{eq:conservation}
\end{equation}
This expresses conservation of kinetic energy: the total kinetic energy stored in the particles remains constant, and only its distribution among the different species changes.
%\rlnote{We should make a note that the equipartition also implies that some of the energy goes into the field energy, so conservation of energy implies that the sum of the final temperature should in fact be a bit lower, as pointed out in Justin's email.}

Second, \cref{eq:evolutionT} yields a very simple result \emph{at thermal equilibrium}. Indeed, once thermal equilibrium is reached, $d T_{\sigma}/dt = 0$ and, \cref{eq:evolutionT} then implies:
\begin{equation} 
w_\sigma T_{\sigma, eq.} = w_{\sigma'}T_{\sigma', eq.} 
\label{eq:equilibriumT}
\end{equation}
This equation predicts that high-weight (resp. low-weight) macroparticles will have an unphysically low (resp. high) temperature at equilibrium, and that the ratio of these temperatures is simply the ratio of the macroparticle weights. In the special case of a plasma consisting of only two species with equal density ($n_1 = n_2$), combining \cref{eq:equilibriumT} with the conservation of energy \cref{eq:conservation} yields the simple predictions for $T_{1,\mathrm{eq.}}, T_{2,\mathrm{eq.}}$ that were given empirically in \cref{sec:collisionlessempirical} (\cref{eq:twospeciesequilibrium1,eq:twospeciesequilibrium2}).

The unphysical equilibrium \cref{eq:equilibriumT} can be reinterpreted by multiplying this equation by $3 k_B/2$ and by using the definition of temperature ($k_B T_\sigma = m_\sigma \langle \boldsymbol{v}^2 \rangle_\sigma / 3$) along with the definition of the macroparticle mass $m_{\mathrm{macro}, \sigma} = w_{\sigma} m_\sigma$:
\begin{equation} 
\frac{1}{2} m_{\mathrm{macro},\sigma} \langle \boldsymbol{v}^2 \rangle_{\sigma} = \frac{1}{2}m_{\mathrm{macro},\sigma'} \langle \boldsymbol{v}^2 \rangle_{\sigma'}
\label{eq:equipartition}
\end{equation}
where, again, $\langle ... \rangle_\sigma$ denotes an average over the population of species $\sigma$. This equation corresponds to equipartition of energy applied to the \emph{macroparticles} rather than to the \emph{physical particles} they represent. In other words, the average kinetic energy of a \emph{macroparticle} is identical across species, regardless of whether its species has low or high weight. As a result, a species with fewer macroparticles (but representing the same total number of physical particles) stores less energy at equilibrium than a species with more macroparticles. As noted in the introduction, there is therefore an unphysical energy transfer from high-weight to low-weight species. This result is also consistent with the equipartition of energy reported in \cite{AngusJCP2022}, where it was shown that, at equilibrium, the energy stored in the particles of a PIC code depends on the total number of \emph{macroparticles} rather than the total number of \emph{physical particles}.

In hindsight, this unphysical equilibrium could have been anticipated from the fact that, at its core, the PIC algorithm sees only a collection of macroparticles (with charge $q_{\mathrm{macro},\sigma}$ and mass $m_{\mathrm{macro},\sigma}$) and retains no information about the fact that these macroparticles represent differing numbers of physical particles (with charge $q_{\sigma}$ and mass $m_{\sigma}$). Indeed, although $w_\sigma$, $q_\sigma$, and $m_\sigma$ may appear explicitly in the source code, every step of the PIC algorithm can ultimately be written in a form that depends only on $q_{\mathrm{macro},\sigma} = w_\sigma q_\sigma$ and $m_{\mathrm{macro},\sigma} = w_\sigma m_\sigma$. For instance, the charge/current deposition effectively uses the product $w_\sigma q_{\sigma} = q_{\mathrm{macro},\sigma}$, and the particle velocity update likewise can be expressed with the ratio $q_{\mathrm{macro},\sigma}/m_{\mathrm{macro},\sigma}$. Thus, since the PIC algorithm effectively carries no information about the underlying physical particles, it cannot recover the physical equipartition $m_\sigma \langle \boldsymbol{v}^2\rangle_\sigma/2 = m_{\sigma'} \langle \boldsymbol{v}^2\rangle_{\sigma'}/2$, and instead produces the unphysical equipartition $m_{\mathrm{macro},\sigma} \langle \boldsymbol{v}^2\rangle_\sigma/2 = m_{\mathrm{macro},\sigma'} \langle \boldsymbol{v}^2\rangle_{\sigma'}/2$.

The considerations of the previous paragraph are quite general, and do not rely on the specifics of the PIC scheme (e.g., field solver, time-stepping scheme, etc.). We therefore expect the unphysical temperature equilibrium \cref{eq:equilibriumT}, and its reinterpretation as unphysical equipartition \cref{eq:equipartition}, to hold across many variants of the PIC algorithm, well beyond the electrostatic, explicit ``energy-conserving'' scheme for which we derived it here. 

These general considerations also suggest that this unphysical equilibrium is essentially unavoidable in the long run. Reaching it, however, takes time, and whether it is actually attained over the course of a given simulation depends on the simulation duration relative to the thermalization timescale. This is where \cref{eq:evolutionT} is instrumental, in that it predicts the exact time evolution of the temperature. These predictions are verified against PIC simulations in the next section.

\subsection{Comparison with PIC simulations over a range of parameters}
\label{sec:comparison}

\begin{figure}
\includegraphics[width=\columnwidth]{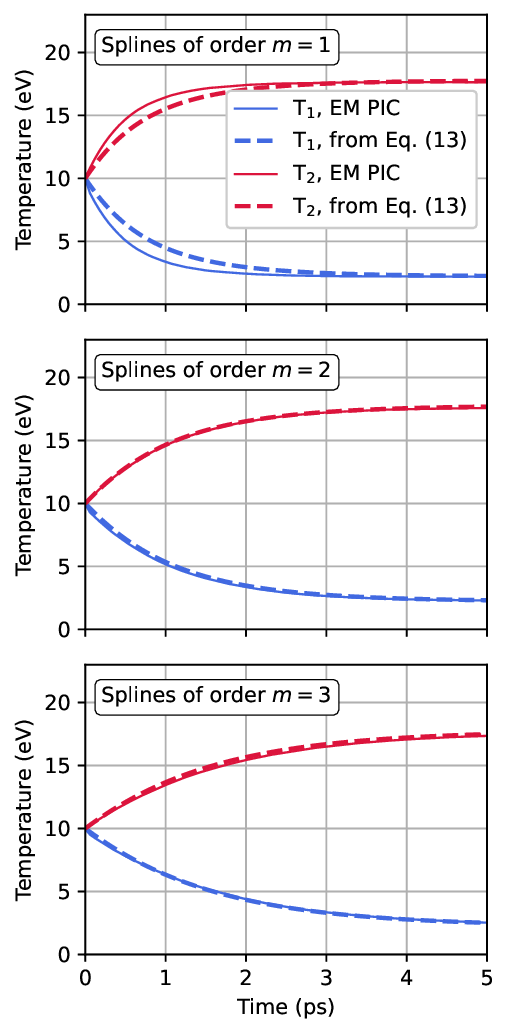}
\caption{Comparison between electromagnetic PIC simulations (EM PIC, solid lines) and the predictions of  \cref{eq:evolutionT,eq:logPIC,eq:dielectricM} (dashed lines), using macroparticle shapes with different spline orders $m$ ($m=1,2,3$, from top to bottom). The PIC simulations use the parameters of \cref{tab:parameters} and case B in \cref{tab:parameters3D}. The blue lines correspond to the electrons (species 1), and the red lines correspond to the ions (species 2).
}
\label{fig:shape_factor}
\end{figure}

In this section, we compare the predictions of \cref{eq:evolutionT,eq:logPIC,eq:dielectricM} (obtained by numerical integration) with the results of PIC simulations, under a range of conditions. All the PIC simulations in this section use the parameters of \cref{tab:parameters}, with the macroparticle weights from case B of \cref{tab:parameters3D}. The weight ratio is therefore fixed at $w_1/w_2 = 8$, which, according to \cref{eq:twospeciesequilibrium1,eq:twospeciesequilibrium2}, also fixes the final equilibrium temperatures $T_{1,\mathrm{eq.}}, T_{2,\mathrm{eq.}}$. Throughout this section, the initial temperatures are chosen to be equal: $T_{1,\mathrm{init.}}=T_{2,\mathrm{init.}}$.

\Cref{eq:evolutionT,eq:logPIC} indicate that the temperature evolution depends on $\log \Lambda_{PIC,\sigma\sigma'}$, which in turn depends on the spline order $m$ used in the macroparticle shape. In order to verify this prediction, we run simulations with different spline orders $m=1, 2, 3$, and display the corresponding temperature evolution in \cref{fig:shape_factor}. We note that, in the literature, these different spline orders are also commonly known as Cloud In Cell (CIC) for $m=1$, Triangular Shaped Cloud (TSC) for $m=2$, and Piecewise Quadratic Cloud Shape (PQS) for $m=3$ \cite{Hockney}.

Overall, the results of the PIC simulations (solid lines in \cref{fig:shape_factor}) show that the relaxation timescale increases with spline order $m$. This is intuitive, since higher spline orders $m$ smooth out the interactions between charged macroparticles and thus reduce the effective collisionality of the PIC algorithm \cite{Okuda,Park,Touati}. Since the temperature evolution here is driven by this effective collisionality, it is natural that this evolution slows down as $m$ increases.

\Cref{fig:shape_factor} also shows that the predictions of \cref{eq:evolutionT} (dashed lines) capture this effect, and that they are in fact in very good agreement with the PIC results, except for $m=1$. Further analysis (see \cref{app:aliases}) reveals that this weaker agreement for $m=1$ is primarily due to the fact that spatial aliases were neglected. The contribution of aliases, which is noticeable for $m=1$, is expected to decrease rapidly for higher $m$ and is thus much less visible for $m=2$ and $m=3$. For the remainder of this section, we use splines of order $m=3$.

Another prediction of \cref{eq:evolutionT,eq:logPIC,eq:dielectricM} is that the relaxation timescale depends on temperature in a complex way, since the temperature appears both in the temperature evolution \cref{eq:evolutionT} and in the dielectric function \cref{eq:dielectricM}. To test this prediction, we run PIC simulations with different initial temperatures: $T_{1,\mathrm{init.}} = T_{2,\mathrm{init.}} =$ 5 eV, 50 eV, 500 eV, 5000 eV. Among other things, this changes the Debye length and how well it is resolved: for the fixed cell size $\Delta x$ used here, we have $\lambda_{D,e} / \Delta x = $ 0.053, 0.17, 0.53, 1.7 respectively. This set of simulations thus spans a range of regimes, from an under-resolved to a marginally resolved Debye length. Note that we are able to run simulations with an under-resolved Debye length without triggering the finite-grid instability only because they use the ``energy-conserving'' gathering scheme \cite{Birdsall}.

\begin{figure}
\includegraphics[width=0.93\columnwidth]{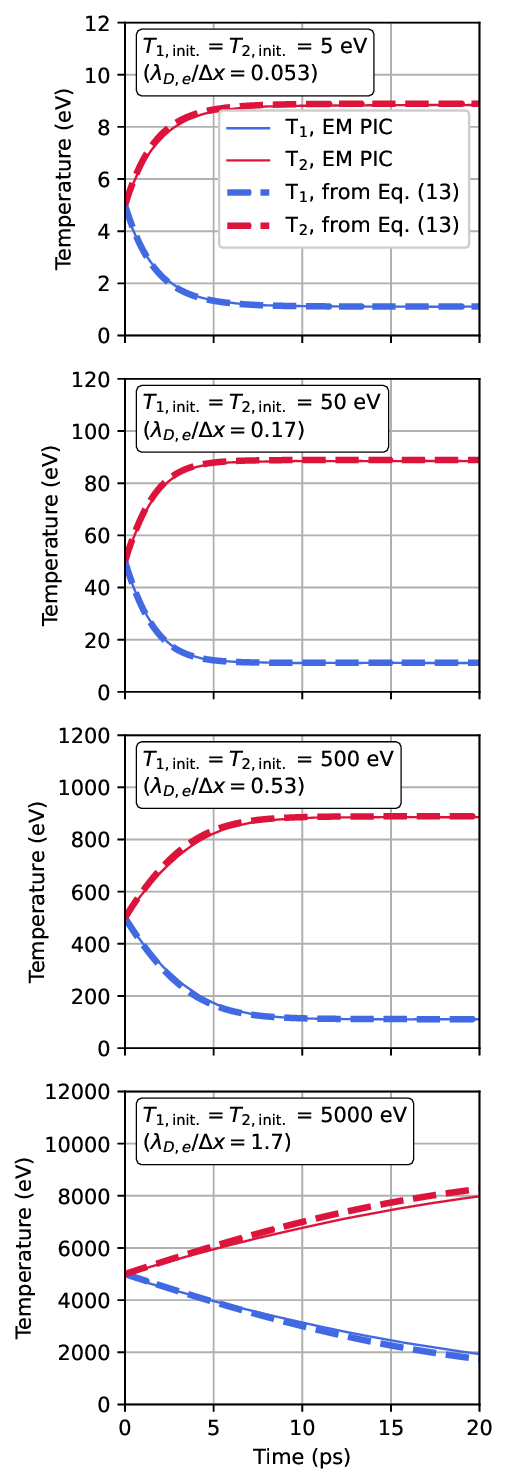}
\caption{Comparison between electromagnetic PIC simulations (EM PIC, solid lines) and the predictions of  \cref{eq:evolutionT,eq:logPIC,eq:dielectricM} (dashed lines), for different initial temperatures ($T_{1,\mathrm{init.}} = T_{2,\mathrm{init.}} =$ 5 eV, 50 eV, 500 eV, 5000 eV, from top to bottom). The PIC simulations use the parameters of \cref{tab:parameters} and case B in \cref{tab:parameters3D}. The blue lines correspond to the electrons (species 1), and the red lines correspond to the ions (species 2).
}
\label{fig:temperature}
\end{figure}

\Cref{fig:temperature} again shows very good agreement between the PIC simulation results (solid lines) and the predictions of \cref{eq:evolutionT,eq:logPIC,eq:dielectricM} (dashed lines), thereby validating these equations over a wide range of parameters. Interestingly, the relaxation timescale depends only weakly on temperature for $\lambda_{D,e} / \Delta x < 1$, but more strongly for $\lambda_{D,e}/\Delta x > 1$. Understanding this behavior motivates a closer look at the behavior of the effective Coulomb logarithm $\log \Lambda_{PIC,\sigma\sigma'}$ as given by \cref{eq:logPIC}.

\section{Approximate expression of $\log \Lambda_{PIC}$ for an electron-ion plasma with similar temperatures $T_e \sim T_i$}
\label{sec:approximate}

In \cref{eq:logPIC}, the expression of $\log \Lambda_{PIC,\sigma \sigma'}$ is given for a plasma consisting of an arbitrary number of species, with arbitrary mass and temperature. However, a very common case is that of a two-species plasma consisting of an electron species and an ion species, with $m_e \ll m_i$. (In this case, we use the notation $\sigma=e, \sigma'=i$ for the electron and ion species respectively and denote the Coulomb logarithm as $\log \Lambda_{PIC,ei}$.) For this case, and when assuming similar temperatures ($T_e \sim T_i$) and equal cell sizes ($\Delta x = \Delta y = \Delta z$), it turns out that $\log \Lambda_{PIC, ei}$ has a much simpler expression. As shown in \cref{app:approximate}, under these conditions, $\log \Lambda_{PIC, ei}$ does not depend on the ion properties $m_i$, $T_i$, $\lambda_{D,i}$ anymore, and instead depends only on $\lambda_{D,e}/\Delta x$. Additionally, it reduces to simple expressions in two asymptotic regimes:
\begin{itemize}
\item Under-resolved Debye length ($\lambda_{D,e}/\Delta x \ll 1$):
\begin{equation}
\log \Lambda_{PIC, ei} \approx 60\times \left(\frac{\lambda_{D,e}}{\Delta x}\right)^4
\label{eq:logPIC_3D_low}
\end{equation}
\item Well-resolved Debye length ($\lambda_{D,e}/\Delta x \gg 1$):
\begin{equation}
\log \Lambda_{PIC, ei} \approx \log \left(\frac{\lambda_{D,e}}{\Delta x}\right)
\label{eq:logPIC_3D_high}
\end{equation}
\end{itemize}
While \cref{eq:logPIC_3D_low,eq:logPIC_3D_high} are specific to 3D simulations, their generalization for 1D and 2D simulations are given in \cref{app:2d1d} (\cref{eq:logPIC_anyD_low,eq:logPIC_anyD_high}).

\begin{figure}
\includegraphics[width=\columnwidth]{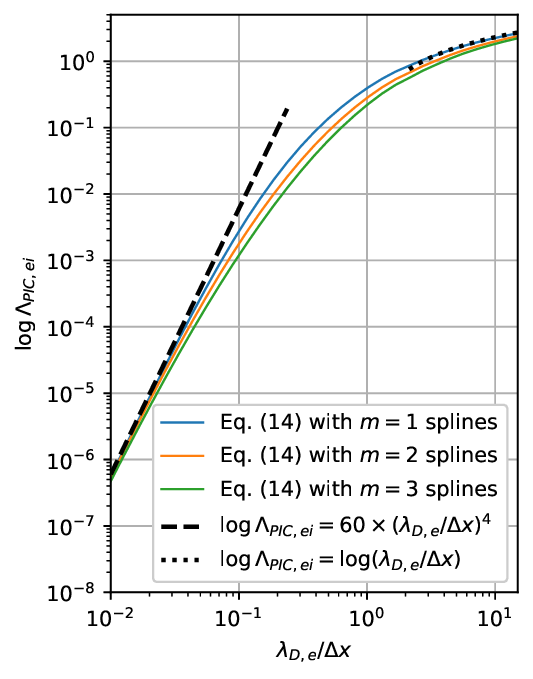}
\caption{Results of the numerical evaluation of the expression for $\log \Lambda_{PIC,ei}$ \cref{eq:logPIC} for an electron-ion plasma and for different spline orders $m$ (solid lines, different colors). The results are compared with the approximate expressions of $\log \Lambda_{PIC,ei}$ \cref{eq:logPIC_3D_low} (dashed line) and \cref{eq:logPIC_3D_high}  (dotted line). While the solid lines were obtained here with $m_i = 2000 \,m_e$, $T_i = T_e$, the results are largely independent of $m_i$ and $T_i$ (as long as $T_i \sim T_e$) as shown in \cref{app:approximate}.
\label{fig:logPIC_3D}}
\end{figure}

\Cref{fig:logPIC_3D} displays $\log \Lambda_{PIC,ei}$, as evaluated numerically from \cref{eq:logPIC}, as a function of $\lambda_{D,e}/\Delta x$ and different spline orders $m$, and compares it with the above approximate expressions \cref{eq:logPIC_3D_low,eq:logPIC_3D_high}. While \cref{fig:logPIC_3D} was obtained with $m_i = 2000 \,m_e$, $T_i = T_e$, the results are largely independent of $m_i$ and $T_i$ (as long as $T_i \sim T_e$) as shown in \cref{app:approximate}.

As can be seen in \cref{fig:logPIC_3D}, the effective Coulomb logarithm for PIC, $\log \Lambda_{PIC,ei}$, varies over several orders of magnitude as $\lambda_{D,e}/\Delta x$ is varied from $10^{-2}$ to $10$. This is in stark contrast with its \emph{physical} counterpart, the Coulomb logarithm that appears in the Spitzer theory for temperature evolution, which typically remains between 2 and 20 across a wide range of plasma conditions.

In addition, \cref{fig:logPIC_3D} confirms that the approximate expressions, $\log \Lambda_{PIC,ei} = 60\times (\lambda_{D,e}/\Delta x)^4$ for $\lambda_{D,e}/\Delta x \ll 1$ and $\log \Lambda_{PIC,ei} = \log(\lambda_{D,e}/\Delta x)$ for $\lambda_{D,e}/\Delta x \gg 1$, agree well with the full expression for $\log \Lambda_{PIC,ei}$ \cref{eq:logPIC} in their respective asymptotic regimes. These asymptotic behaviors can be understood intuitively, and stem from the fact that collisions between charged macroparticles are suppressed by Debye shielding for impact parameters larger than $\lambda_{D,e}$, and by the smoothing effect of the grid and macroparticle shape for impact parameters smaller than $\Delta x$:
\begin{itemize}
\item For $\lambda_{D,e}/\Delta x \gg 1$, these two ranges of impact parameters do not overlap, leaving a window of unsuppressed collisions. The effective Coulomb logarithm then behaves like the physical one, $\log(\lambda_{D,e}/b_{\mathrm{min}})$, but with the lower cut-off distance $b_{\mathrm{min}}$ replaced by the cell size $\Delta x$.
\item For $\lambda_{D,e}/\Delta x \ll 1$, the two ranges overlap, so collisions are suppressed at all impact parameters, and the effective PIC collisionality is strongly reduced overall.
\end{itemize}
We note that similar asymptotic regimes were also discussed in \cite{Okuda,Langdon} when calculating the scattering cross-section and collision frequency between finite-size charged particles.

Finally, \cref{fig:logPIC_3D} also shows that $\log \Lambda_{PIC,ei}$ decreases with increasing spline order $m$, in agreement with the observations of \cref{fig:shape_factor}. Interestingly, in the regimes $\lambda_{D,e}/\Delta x \ll 1$ and  $\lambda_{D,e}/\Delta x \gg 1$, the influence of the spline order appears to vanish, and indeed the parameter $m$ does not enter in the approximate formula \cref{eq:logPIC_3D_low,eq:logPIC_3D_high}.

Overall, the above discussion of $\log \Lambda_{PIC,ei}$ can be used to obtain an estimate of the relaxation timescale, in the case of an electron-ion plasma with $T_e \sim T_i$. For a two-species plasma with $n_e = n_i = n$, energy conservation \cref{eq:conservation} (which yields $T_e + T_i = T_{e,\mathrm{init.}} + T_{i,\mathrm{init.}}$) can be used to recast the temperature evolution \cref{eq:evolutionT} as an equation for $\Delta T_e = T_e - T_{e,\mathrm{eq.}}$, where $T_{e,\mathrm{eq.}}$ is given by \cref{eq:twospeciesequilibrium1}:
\begin{equation}
\frac{d \,\Delta T_e}{dt} = -\frac{1}{\tau_{ei}}\Delta T_{e}
\end{equation}
with:
\begin{equation}
\tau_{ei} = \frac{3(2\pi)^{3/2} n}{2\;\omega_{p,e}^2 \omega_{p,i}^2} \frac{(k_B T_e/m_e)^{3/2}}{ (w_{e} + w_{i}) \log \Lambda_{PIC, e i}}
\label{eq:relaxationtimescale}
\end{equation}
where we used the fact that $T_e/m_e \gg T_i/m_i$ to simplify the expression of $\tau_{ei}$. 

The above expression for the relaxation timescale $\tau_{ei}$ reveals a number of important properties. Since $\omega_{p,i}^2 \propto 1/m_i$, the relaxation timescale is simply proportional to $m_i$ ($\tau_{ei}\propto m_i$): heavier ions lead to slower relaxation. The dependence on the macroparticle weights is likewise simple, with $\tau_{ei}$ inversely proportional to the sum of the species' weights $w_e + w_i$. Finally, this expression explains the weak temperature dependence observed in \cref{fig:temperature} for $\lambda_{D,e}/\Delta x < 1$: in this regime, the fact that the numerator $(k_B T_e/m_e)^{3/2}$ increases with temperature is compensated by the fact that $\log \Lambda_{PIC,ei}$ in the denominator rapidly increases with $\lambda_{D,e}$ (see \cref{fig:logPIC_3D}) which also increases with temperature.

\Cref{eq:relaxationtimescale} also suggests strategies for minimizing the impact of the unphysical energy transfer between macroparticles of different weights: the aim is to make the relaxation timescale $\tau_{ei}$ much longer than that of the simulation. This can be done either by reducing the macroparticle weights $w_e, w_i$, or by reducing $\log \Lambda_{PIC,ei}$, which, according to \cref{fig:logPIC_3D}, means increasing the spline order $m$ or increasing the cell size $\Delta x$. All of these strategies ultimately amount to reducing the effective collisionality of the PIC algorithm.
Increasing $\Delta x$ is especially effective in the regime $\lambda_{D,e}/\Delta x \ll 1$, where $\log \Lambda_{PIC,ei} \propto 1/\Delta x^4$ and hence $\tau_{ei} \propto \Delta x^4$. Since this is the regime in which the Debye length is under-resolved, it is only a viable option if the physics at the scale of $\lambda_{D,e}$ is unimportant for the effects under study in the simulation, and if the PIC scheme remains numerically stable for $\lambda_{D,e}/\Delta x < 1$, as is the case for the explicit ``energy-conserving'' scheme.

\section{Implications of this work and conclusion}

In summary, we have shown that whenever macroparticles have different weights, the PIC algorithm produces a systematic, unphysical transfer of energy from the higher-weight to the lower-weight macroparticles. As a result, rather than relaxing towards the expected physical equilibrium $T_{\sigma,\mathrm{eq.}} = T_{\sigma',\mathrm{eq.}}$ (where $\sigma$ and $\sigma'$ label species with different weights), the system relaxes towards the unphysical state $w_\sigma T_{\sigma,\mathrm{eq.}} = w_{\sigma'} T_{\sigma',\mathrm{eq.}}$. This state can be reinterpreted as equipartition \emph{applied to the macroparticles themselves}, and it is intrinsic to the PIC algorithm: at its core, the algorithm only ever sees macroparticles, and carries no information about the physical particles they represent. Accordingly, the effect is observed both for electrostatic and electromagnetic PIC, and in 3D as well as (for the temperature along the grid axes) in 1D and 2D.

Reaching this unphysical equilibrium nevertheless takes time. The dynamics is governed by the effective collisionality of the PIC algorithm, that is, by the fact that the algorithm naturally captures Coulomb collisions between charged macroparticles, albeit in a distorted form. We derived an equation for the resulting temperature evolution (\cref{eq:evolutionT} for 3D PIC, and \cref{eq:evolutionT_anyD} for its generalization to 1D and 2D) whose predictions agree well with PIC simulations over a wide range of parameters. A notable feature of this description is that the effective Coulomb logarithm $\log \Lambda_{PIC,\sigma\sigma'}$ can vary over several orders of magnitude with $\lambda_{D,e}/\Delta x$, in stark contrast with its physical counterpart.

The main implication of this work is that PIC simulations using macroparticles of different weights require more care than was perhaps previously thought. This is particularly true when the physics under study in the simulation depends on obtaining the correct temperatures. The temperatures measured in such simulations may be wrong either because the unphysical equilibrium has been reached, or, in runs too short for that, because the unphysical energy transfer has already caused the temperatures to deviate from their physical evolution. Our equation for the temperature evolution provides a way to estimate the magnitude of this effect in advance. Alternatively, one can gauge the importance of this effect by verifying the convergence of the results with respect to the macroparticle weights. When doing so, it is essential to vary the weight of the low-weight and high-weight macroparticles \emph{independently}, since changing both while keeping their ratio fixed leaves the unphysical equilibrium unchanged and can lead to a false sense of convergence.

Where the effect is found to matter, it can be mitigated by lengthening the associated timescale, i.e. by reducing the effective collisionality of the algorithm. This can be done by uniformly reducing the macroparticle weights (that is, by using more macroparticles), by using higher-order splines, or by deliberately under-resolving the Debye length. This last option is viable only when Debye-scale physics is unimportant for the problem at hand. While the guidelines in this paragraph were derived specifically for a two-species electron-ion plasma (where compact expressions are available), we expect them to hold more generally.

Our work opens up several lines of research beyond the present paper:

\paragraph{Different weights within a single species:} This paper focused on the case where weights differ between species but remain uniform within each species. However, the same unphysical energy transfer between low- and high-weight macroparticles should also occur whenever weights vary \emph{within} a species. This is immediately apparent when the weights take a discrete set of values: the macroparticles of each weight can simply be relabeled as distinct ``species'', to which our analysis (including \cref{eq:evolutionT}) applies directly. By contrast, the case where the distribution of weights is continuous and/or evolves dynamically (see some of the examples in the introduction) will require a new theoretical framework. Note that, when the species with varying weights is an electron species, the unphysical effect could be especially pronounced, since the relaxation timescale would no longer involve the ion mass (as it does in \cref{sec:approximate}) and would likely be considerably shorter.

\paragraph{Coupling with a Monte Carlo collision module:} This paper deals with the bare PIC algorithm, \emph{in the absence} of any Monte Carlo collision (MCC) module \cite{TA77,Nanbu1997,Nanbu1998}. Monte Carlo collisions are expected to drive the system towards the physical equilibrium $T_{\sigma,\mathrm{eq.}} = T_{\sigma',\mathrm{eq.}}$, even for macroparticles of unequal weights, since modern weight-aware MCC algorithms \cite{Nanbu1998,Sentoku2008,Perez2012,Higginson2020,Angus_collisions} (unlike the PIC algorithm itself, as demonstrated in this paper) do carry information about the underlying physical particles. In coupled PIC-MCC simulations, one should therefore expect a competition between the MCC algorithm, which drives the system towards the physical equilibrium $T_{\sigma,\mathrm{eq.}} = T_{\sigma',\mathrm{eq.}}$, and the PIC algorithm, which drives it towards the unphysical equilibrium $w_\sigma T_{\sigma,\mathrm{eq.}} = w_{\sigma'}T_{\sigma',\mathrm{eq.}}$. We have indeed observed this competition in separate PIC-MCC simulations, and will publish the results and the corresponding theoretical framework in a follow-up paper.

\begin{acknowledgments}
We thank Jaehong Park for sharing insightful observations which prompted further investigation into this topic. This research used the open-source PIC code WarpX \cite{WarpX}. Primary WarpX contributors are with LBNL, LLNL, CEA-LIDYL, SLAC, DESY, CERN, Helion Energy, TAE Technologies and Realta Fusion. We acknowledge all WarpX contributors. This research also made use of PlasmaPy version 2026.2.0, a community-developed open source Python package for plasma research and education \cite{plasmapy_community_2026_18706665}. 

% SciDAC-5 KISMET
This material is based upon work supported by the KISMET collaboration, a project of the U.S. Department of Energy, Office of Science, Office of Advanced Scientific Computing Research and Office of High Energy Physics, Scientific Discovery through Advanced Computing (SciDAC) program. This research used resources of the National Energy Research Scientific Computing Center, a DOE Office of Science User Facility supported by the Office of Science of the U.S. Department of Energy under Contract No. DE-AC02-05CH11231 using NERSC award FES-ERCAP0027617.
\end{acknowledgments}

\appendix

\section{Generalization for 1D and 2D PIC simulations}
\label{app:2d1d}

While the main text focuses on 3D simulations, in this appendix we discuss how these results extend to 1D and 2D simulations. 

In this context, the terms 1D, 2D and 3D refer specifically to the spatial grid. The macroparticle velocity $\boldsymbol{v}$, on the other hand, always remains a 3D vector. (This configuration is also known as 1D3V, 2D3V, or 3D3V PIC.) In the text below, $D=1, 2, 3$ denotes the dimensionality of the spatial grid, and we adopt the standard labeling convention for the grid axes: $x$ in 1D, $x, y$ in 2D, and $x, y, z$ in 3D. (Readers that are familiar with WarpX might notice that this differs from the labeling convention used internally in the code.)

As we show below, a key difference in 1D and 2D is that the velocity components along the grid axes ($v_x$ in 1D, $v_x, v_y$ in 2D) behave differently from the components along the invariant dimensions ($v_y, v_z$ in 1D, $v_z$ in 2D). We therefore distinguish the temperature $T_{\sigma,\parallel}$ along the grid axes, defined by
\begin{align}
k_B T_{\sigma,\parallel} &\equiv \left\{ \begin{array}{l l}
m_\sigma\langle v_x^2\rangle_\sigma & \mathrm{for}\,D=1 \\[6pt]
\frac{m_\sigma}{2}\langle v_x^2 + v_y^2 \rangle_\sigma & \mathrm{for}\,D=2\\[6pt]
\frac{m_\sigma}{3}\langle v_x^2 + v_y^2 + v_z^2\rangle_\sigma & \mathrm{for}\,D=3
\end{array}\right.
\label{eq:gridalignedT}
\end{align}
and the temperature $T_{\sigma,\perp}$ along the invariant dimensions
\begin{align}
k_B T_{\sigma,\perp} &\equiv \left\{ \begin{array}{l l}
\frac{m_\sigma}{2}\langle v_y^2 + v_z^2\rangle_\sigma & \mathrm{for}\,D=1 \\[6pt]
m_\sigma\langle v_z^2 \rangle_\sigma & \mathrm{for}\,D=2\\[6pt]
\mathrm{undefined} & \mathrm{for}\,D=3
\end{array}\right.
\label{eq:gridinvariantT}
\end{align}

Another important point when discussing 1D, 2D and 3D results is that the dimensionality of the weights $w_\sigma$ depends on $D$: $w_\sigma$ is a (dimensionless) number of physical particles for $D=3$, a linear density of physical particles for $D=2$, or an areal density of physical particles for $D=1$. The quantity $w_\sigma \Delta x^{3-D}$, however, is always dimensionless. %(Note that this definition of the weight in 1D and 2D differs from that of \cite{Park}.)

\subsection{Observation of an unphysical temperature equilibrium}

\begin{figure*}
\includegraphics[width=\textwidth]{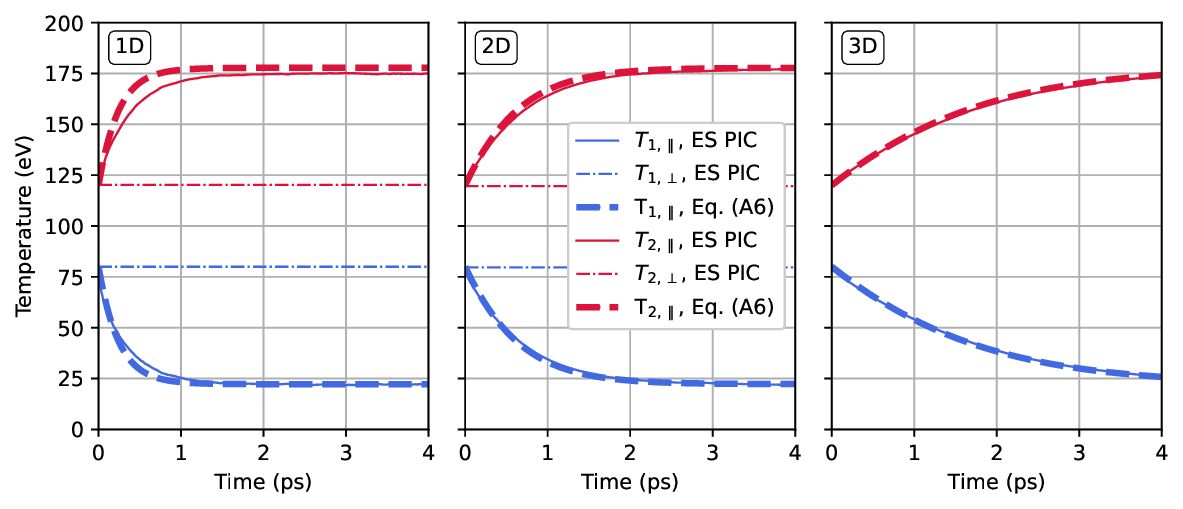}
\caption{Evolution of the temperature of each species ($T_1$ in blue and $T_2$ in red) in electrostatic PIC simulations of a uniform plasma with parameters given in \cref{tab:parameters,tab:parameters123D} in 1D (left), 2D (center), and 3D (right). The solid and dot-dashed curves represent the temperature along the grid axes $T_{\sigma,\parallel}$ and along the invariant direction $T_{\sigma,\perp}$ respectively (see \cref{eq:gridalignedT,eq:gridinvariantT}) obtained from PIC simulations. The dashed line corresponds to the prediction of \cref{eq:evolutionT_anyD}. \label{fig:dims}}
\end{figure*}

\Cref{fig:dims} shows the results of 1D, 2D and 3D electrostatic PIC simulations of a two-species uniform plasma using the same parameters as in \cref{tab:parameters} (keeping the same cell size, as well as the same total number of cells, i.e. 4096 cells in 1D, 64$\times$ 64 cells in 2D, $16\times16 \times 16$ cells in 3D), and using a fixed ratio of weights between species $w_1/w_2 = 8$, as detailed in \cref{tab:parameters123D}. The plasma species are initialized with $k_B T_{1,\textrm{init.}} = 80\,\mathrm{eV}$, $k_B T_{2,\mathrm{init.}} = 120\,\mathrm{eV}$.

As can be seen in \cref{fig:dims}, the temperature along the invariant dimensions, $T_{\sigma,\perp}$, remains constant. This is in fact expected: as discussed in the main text, the velocity distributions evolve because of the effective collisions between charged macroparticles, which, in the PIC algorithm, are mediated by the electric field on the grid. In 1D and 2D  electrostatic PIC, the electric field components along the invariant dimensions ($y$,$z$ in 1D, $z$ in 2D) are zero, and thus the corresponding components of the macroparticle velocities cannot change.

On the other hand, \cref{fig:dims} also shows that the temperature along the grid axes, $T_{\sigma,\parallel}$, evolves towards the same unphysical equilibrium in 1D, 2D and 3D, i.e.:
\begin{align} 
T_{1,\parallel,\mathrm{eq.}} &= \frac{1}{1 + w_1/w_2} (T_{1,\mathrm{init.}} + T_{2,\mathrm{init.}}) \label{eq:twospeciesequilibrium1_anyD} \\
T_{2,\parallel,\mathrm{eq.}} &= \frac{1}{1 + w_2/w_1} (T_{1,\mathrm{init.}} + T_{2,\mathrm{init.}}) 
\label{eq:twospeciesequilibrium2_anyD}
\end{align}
This confirms that the unphysical thermal equilibrium described in the main text also exists in 1D and 2D.

\begin{table*}
\begin{center}
  \begin{tabular}{ccc}
    \hline
    Case & Species 1 & Species 2\\
    \hline
    3D & 8 macroparticles/cell ($w_1 = 191 $) & 64 macroparticles/cell ($w_2 = 23.8$)  \\
    2D & 8 macroparticles/cell ($w_1\Delta x = 191$) & 64 macroparticles/cell ($w_2\Delta x = 23.8$)  \\
    1D & 8 macroparticles/cell ($w_1\Delta x^2 = 191$) & 64 macroparticles/cell ($w_2\Delta x^2 = 23.8$)\\
    \hline
  \end{tabular}  
\end{center}  
\caption{Number of macroparticles (and corresponding weights) for the simulations displayed in \cref{fig:dims}.}
\label{tab:parameters123D} 
\end{table*}

\subsection{Equation for the temperature evolution}

As shown in \cref{app:derivationT}, the equations for the temperature evolution, which are given for 3D in the main text (\cref{eq:evolutionT,eq:logPIC,eq:dielectricM}), can be generalized for $D=1,2,3$:
\begin{widetext}
\begin{equation}
\frac{d\, T_{\sigma,\perp}}{dt} =0
\end{equation}
\begin{equation}
\frac{d\, T_{\sigma,\parallel}}{dt} =\sum_{{\sigma'}}  \frac{2\;\omega_{p,\sigma}^2 \omega_{p,{\sigma'}}^2 \Delta x^{3-D}}{D(2\pi)^{3/2} n_\sigma} \frac{ w_{\sigma'} T_{\sigma',\parallel} - w_\sigma T_{\sigma,\parallel} }{\left[ \, k_B T_{\sigma,\parallel}/m_\sigma + k_B T_{\sigma',\parallel}/m_{\sigma'}\,\right]^{3/2}}\log \Lambda_{PIC, \sigma \sigma'}
\label{eq:evolutionT_anyD}
\end{equation}
\begin{equation}
\log \Lambda_{PIC, \sigma \sigma'} = \int_g\frac{d^D\boldsymbol{k}\;\Delta x^{D-3}}{2^{D-2}\pi^{D-3/2}}\frac{S_m^4(\boldsymbol{k}) k}{K^4(\boldsymbol{k})} \int_{-\infty}^\infty \!\!\!\!\!d\alpha\frac{\alpha^2e^{-\alpha^2}}{\left|\epsilon\left(\boldsymbol{k}, \frac{\alpha k}{\sqrt{ \frac{m_\sigma}{2k_B T_{\sigma,\parallel}} + \frac{m_{\sigma'}}{2k_B T_{\sigma',\parallel}} }} \right)\right|^2}
\label{eq:logPIC_anyD}
\end{equation}
\begin{equation} 
\epsilon(\boldsymbol{k}, \omega) = 1 - \sum_{\sigma''} \frac{S_m^2(\boldsymbol{k})}{2K^2(\boldsymbol{k}) \lambda_{D,\sigma''}^2}  Z'\left( \frac{\omega}{k} \sqrt{\frac{m_{\sigma''}}{2 k_B T_{\sigma'',\parallel}}}\,  \right)
\label{eq:dielectricM_anyD} 
\end{equation}
\end{widetext}
where $\lambda_{D,\sigma''} = \sqrt{k_B T_{\sigma'',\parallel}/m_{\sigma''}\omega_{p, \sigma''}^2}$ is the Debye length associated with species $\sigma''$. It is worth pointing out that, in \cref{eq:logPIC_anyD}, $\log \Lambda_{PIC, \sigma \sigma'}$ is always dimensionless, regardless of $D$.

In \cref{fig:dims}, \cref{eq:evolutionT_anyD,eq:logPIC_anyD,eq:dielectricM_anyD} are integrated numerically and compared with the results of the PIC simulations. As the figure shows, the agreement between these predictions and the PIC results is very good, except perhaps in 1D. A likely reason is that the assumption used in \cref{app:derivationT}'s derivation (namely that the velocity distribution remains Maxwellian) may hold less well in 1D than in 2D and 3D. Indeed, it is known that the effect of intra-species collisions, which would normally keep a given species' velocity distribution Maxwellian, are strongly suppressed in 1D PIC, due to a phenomenon known as \emph{kinetic blocking} \cite{Eldridge,Fouvry}. Despite this somewhat weaker agreement in 1D, \cref{eq:evolutionT_anyD,eq:logPIC_anyD,eq:dielectricM_anyD} still correctly predict the relative timescales of the 1D, 2D, and 3D evolution, as well as the final equilibrium.

\subsection{Approximate expression of $\log \Lambda_{PIC}$ for an electron-ion plasma with similar temperatures $T_e \sim T_i$}

In the main text, compact expressions for $\log \Lambda_{PIC,\sigma\sigma'}$ were given in 3D, in the specific case of a two-species electron-ion plasma with similar temperatures $T_e \sim T_i$ and with equal cell sizes ($\Delta x = \Delta y = \Delta z$). These compact expressions were given in the form of asymptotic formulas for $\lambda_{D,e}/\Delta x \ll 1$ and $\lambda_{D,e}/\Delta x \gg 1$. As shown in \cref{app:approximate}, these asymptotic formulas generalize for $D=1,2,3$: 

\begin{itemize}
\item for $\lambda_{D,e}/\Delta x \ll 1$:
\begin{align}
\log \Lambda_{PIC, ei} &\approx \left\{ \begin{array}{l l}
31 \times \left( \frac{\lambda_{D,e}}{\Delta x}\right)^4 &\mathrm{for}\,D=1\\ [8pt]
47 \times \left( \frac{\lambda_{D,e}}{\Delta x}\right)^4 &\mathrm{for}\,D=2\\[8pt]
60 \times \left( \frac{\lambda_{D,e}}{\Delta x}\right)^4 &\mathrm{for}\,D=3
\end{array}\right.
\label{eq:logPIC_anyD_low}
\end{align}
\item for $\lambda_{D,e}/\Delta x \gg 1$:
\begin{equation}
\log\Lambda_{PIC,ei} \approx \left\{ \begin{array}{l l}
\pi \left(\frac{\lambda_{D,e}}{\Delta x}\right)^2  &\mathrm{for}\,D=1\\ [8pt]
\frac{\pi^2}{4} \left(\frac{\lambda_{D,e}}{\Delta x}\right)  &\mathrm{for}\,D=2\\[8pt]
\log \left(\frac{\lambda_{D,e}}{\Delta x}\right) &\mathrm{for}\,D=3
\end{array}\right.
\label{eq:logPIC_anyD_high}
\end{equation}
\end{itemize}

Thus, for $\lambda_{D,e}/\Delta x \ll 1$, one has $\log\Lambda_{PIC,ei} \propto (\lambda_{D,e}/\Delta x)^4$ in any dimension $D$. The possibility of drastically suppressing the effective collisionality of PIC by deliberately under-resolving the Debye length, discussed in the main text for 3D, therefore carries over to 1D and 2D.

\begin{figure*}
\includegraphics[width=\linewidth]{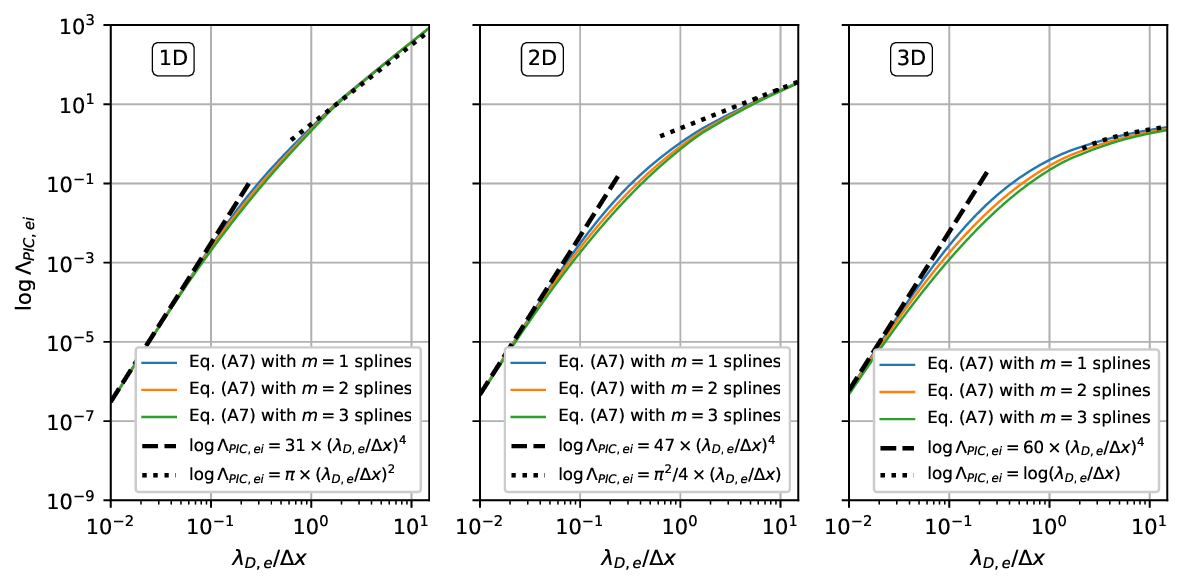}
\caption{Results of the numerical evaluation of the expression for $\log \Lambda_{PIC,ei}$ \cref{eq:logPIC_anyD} for an electron-ion plasma and for different spline orders $m$ (solid lines, different colors). The results are compared with the approximate expressions of $\log \Lambda_{PIC,ei}$ \cref{eq:logPIC_anyD_low} (dashed line) and \cref{eq:logPIC_anyD_high}  (dotted line). While the solid lines were obtained here with $m_i = 2000 \,m_e$, $T_i = T_e$, the results are largely independent of $m_i$ and $T_i$ (as long as $T_i \sim T_e$) as shown in \cref{app:approximate}.
\label{fig:logPIC_any}}
\end{figure*}

The approximate expressions of $\log \Lambda_{PIC,ei}$ in \cref{eq:logPIC_anyD_low} (for $\lambda_{D,e}/\Delta x \ll 1$) and \cref{eq:logPIC_anyD_high}  (for $\lambda_{D,e}/\Delta x \gg 1$) are compared with the numerical evaluation of the original expression for $\log \Lambda_{PIC}$ \cref{eq:logPIC_anyD}. While \cref{fig:logPIC_any} was obtained with $m_i = 2000 \,m_e$, $T_i = T_e$, the results are largely independent of $m_i$ and $T_i$ (as long as $T_i \sim T_e$) as shown in \cref{app:approximate}. Overall, this figure shows good agreement between the asymptotic formulas \cref{eq:logPIC_anyD_low,eq:logPIC_anyD_high} and the full expression of  $\log \Lambda_{PIC,ei}$.

\section{Derivation of the equation for the temperature evolution}
\label{app:derivationT}

In the main text, the equations for the temperature evolution are given specifically in 3D and under the assumption that spatial aliases are negligible (see \cref{eq:evolutionT,eq:logPIC,eq:dielectricM}). In this appendix, we provide a derivation of these equations, and do so in a more general setting: the derivation is formulated so as to apply to 1D, 2D, and 3D, and the effect of spatial aliases is retained throughout. Specializing the final results of this appendix to 3D and neglecting spatial aliases then yields \cref{eq:evolutionT,eq:logPIC,eq:dielectricM}.

In the following, and as in \cref{app:2d1d}, $D=1,2,3$ denotes the dimensionality of the spatial grid. In the PIC simulations, the macroparticle velocity is always a 3D vector, but, as shown in \cref{app:2d1d}, in 1D and 2D the velocity components along the invariant dimensions of the grid ($v_y, v_z$ in 1D; $v_z$ in 2D) do not evolve in time. We therefore need only study the evolution of the velocity components along the grid axes, namely $v_x$ in 1D, $v_x, v_y$ in 2D, and $v_x, v_y, v_z$ in 3D. Accordingly, throughout the rest of this appendix, $\boldsymbol{v}$ denotes a $D$-dimensional vector corresponding to the velocity components along the grid axes, and $f_\sigma(\boldsymbol{v}, t)$ denotes the distribution function over this $D$-dimensional velocity space.

As discussed in \cite{Touati,Park}, the equations for the evolution of the distribution function $f_\sigma(\boldsymbol{v}, t)$, which were given for 3D in the main text in \cref{eq:BGLequation,eq:BGLoperator1,eq:BGLoperator2,eq:dielectric}, can be written more generally for a $D$-dimensional space and read:
\begin{widetext}
\begin{align}
\frac{\partial\, f_\sigma(\boldsymbol{v},t)}{\partial t} &= - \sum_{{\sigma'}}\frac{\partial \,}{\partial \boldsymbol{v}}\cdot \int d^D\boldsymbol{v}'\; \left[\frac{m_{\sigma'}}{m_\sigma} w_{\sigma'} \boldsymbol{Q}_{\sigma {\sigma'}}\cdot\frac{\partial \;}{\partial \boldsymbol{v}} - w_\sigma \boldsymbol{Q'}_{\sigma {\sigma'}}\cdot\frac{\partial \;}{\partial \boldsymbol{v'}} \right] f_\sigma(\boldsymbol{v},t)f_{\sigma'}(\boldsymbol{v'},t)
\label{eq:BGLequation_anyD}\\
\boldsymbol{Q}_{\sigma {\sigma'}} &= - \pi \frac{\omega_{p,\sigma}^2 \omega_{p,{\sigma'}}^2}{n_\sigma}\int_g\frac{d^D\boldsymbol{k}}{(2\pi)^D}\sum_{\boldsymbol{p} \boldsymbol{p'}}\frac{S_m^2(\boldsymbol{k}_{\boldsymbol{p}})S_m^2(\boldsymbol{k}_{\boldsymbol{p'}})}{|\epsilon(\boldsymbol{k}, \boldsymbol{k}_{\boldsymbol{p}}\cdot\boldsymbol{v})|^2}\frac{\boldsymbol{k}_{\boldsymbol{p}} \,\boldsymbol{k}_{\boldsymbol{p}}}{K^4(\boldsymbol{k})}
\delta(\boldsymbol{k}_{\boldsymbol{p}}\cdot\boldsymbol{v}-\boldsymbol{k}_{\boldsymbol{p'}}\cdot\boldsymbol{v'}, 2\pi/\Delta t ) \label{eq:BGLoperator1_anyD}
\\
\boldsymbol{Q'}_{\sigma {\sigma'}} &= - \pi \frac{\omega_{p,\sigma}^2 \omega_{p,{\sigma'}}^2}{n_\sigma}\int_g\frac{d^D\boldsymbol{k}}{(2\pi)^D}\sum_{\boldsymbol{p} \boldsymbol{p'}}\frac{S_m^2(\boldsymbol{k}_{\boldsymbol{p}})S_m^2(\boldsymbol{k}_{\boldsymbol{p'}})}{|\epsilon(\boldsymbol{k}, \boldsymbol{k}_{\boldsymbol{p}}\cdot\boldsymbol{v})|^2}\frac{\boldsymbol{k}_{\boldsymbol{p}}\,\boldsymbol{k}_{\boldsymbol{p'}}}{K^4(\boldsymbol{k})}
\delta(\boldsymbol{k}_{\boldsymbol{p}}\cdot\boldsymbol{v}-\boldsymbol{k}_{\boldsymbol{p'}}\cdot\boldsymbol{v'}, 2\pi/\Delta t )
\label{eq:BGLoperator2_anyD} \\
\epsilon( \boldsymbol{k}, \omega) &= 1 + \sum_{\sigma''} \frac{\omega_{p,\sigma''}^2}{K^2(\boldsymbol{k})}\sum_{\boldsymbol{p''}} S_m^2(\boldsymbol{k}_{\boldsymbol{p''}})\int d^D\boldsymbol{v}'' \;\boldsymbol{k}_{\boldsymbol{p''}}\cdot\frac{\partial f_{\sigma''}(\boldsymbol{v}'',t)}{\partial \boldsymbol{v}''}\frac{\Delta t}{2}\cot\left[ (\omega - \boldsymbol{k}_{\boldsymbol{p''}}\cdot \boldsymbol{v}'')\frac{\Delta t}{2}\right]
\label{eq:dielectric_anyD}
\end{align}
\end{widetext}
where the integrals over $\boldsymbol{k}$ are integral over the $D$-dimensional first Brillouin zone. The expressions of $S_m(\boldsymbol{k})$ (i.e., the Fourier transform of the shape factor, when using splines of order $m$) and $K(\boldsymbol{k})$ (i.e., the Fourier transform of the discrete Laplacian operator) were given in the main text for 3D as $S_m(\boldsymbol{k}) = [\,\mathrm{sinc}(k_x\Delta x/2)\,\mathrm{sinc}(k_y\Delta y/2)\,\mathrm{sinc}(k_z\Delta z/2)\,]^{m+1}$ and $K^2(\boldsymbol{k}) = k_x^2 \mathrm{sinc}(k_x\Delta x/2)^2 + k_y^2 \mathrm{sinc}(k_y\Delta y/2)^2 + k_z^2 \mathrm{sinc}(k_z\Delta z/2)^2$. These expressions remain valid in 2D by setting $k_z = 0$, and in 1D by setting $k_y=0, k_z = 0$.

From the above equations, one can derive the equation for the evolution of the temperature along the axes of the grid (i.e., $x$ in 1D; $x, y$ in 2D; $x,y,z$ in 3D) from its definition as an integral involving the distribution function $ f_\sigma(\boldsymbol{v},t)$:
\begin{equation} 
k_B T_{\sigma, \parallel}(t) = \frac{m_\sigma}{D} \int d^D\boldsymbol{v}\;  f_\sigma(\boldsymbol{v},t)\,\boldsymbol{v}^2 
\label{eq:def_temperature_anyD}
\end{equation}
Note that this equation matches the definition \cref{eq:gridalignedT} given in \cref{app:2d1d}. In 3D, $T_{\sigma,\parallel}$ is simply the temperature $T_{\sigma}$, and the above equation matches \cref{eq:def_temperature} given in the main text.

To make this derivation tractable, we make two simplifying assumptions:
\begin{itemize}
\item We assume that the velocity distribution remains Maxwellian at all time, with a time-dependent temperature:
\begin{equation}
f_{\sigma}(\boldsymbol{v}, t) = \left(\frac{m_{\sigma}}{2\pi k_B T_{\sigma,\parallel}(t)}\right)^{D/2}\!\!\!\!\exp\left( -\frac{m_{\sigma}\boldsymbol{v}^2}{2k_B T_{\sigma,\parallel}(t)}\right)
\label{eq:Maxwellian_anyD}
\end{equation}
\item We neglect time aliases (but as mentioned earlier, we retain spatial aliases until the end of this appendix). This implies that $\delta(\boldsymbol{k}_{\boldsymbol{p}}\cdot\boldsymbol{v}-\boldsymbol{k}_{\boldsymbol{p'}}\cdot\boldsymbol{v'}, 2\pi/\Delta t )$ (defined in \cref{sec:evolution-distribution} as a sum over time aliases) reduces to the ordinary Dirac delta function $\delta(\boldsymbol{k}_{\boldsymbol{p}}\cdot\boldsymbol{v}-\boldsymbol{k}_{\boldsymbol{p'}}\cdot\boldsymbol{v'})$.
\end{itemize}

Combining \cref{eq:def_temperature_anyD} and \cref{eq:BGLequation_anyD}, we have:
\begin{widetext}
\begin{align}
\frac{d\,k_BT_{\sigma, \parallel}}{dt} &= - \sum_{{\sigma'}} \frac{m_\sigma}{D} \int d^D\boldsymbol{v}\; \boldsymbol{v}^2\frac{\partial \,}{\partial \boldsymbol{v}}\cdot \int d^D\boldsymbol{v}'\; \left[\frac{m_{\sigma'}}{m_\sigma} w_{\sigma'} \boldsymbol{Q}_{\sigma {\sigma'}}\cdot\frac{\partial \;}{\partial \boldsymbol{v}} - w_\sigma \boldsymbol{Q'}_{\sigma {\sigma'}}\cdot\frac{\partial \;}{\partial \boldsymbol{v'}} \right] f_\sigma(\boldsymbol{v},t)f_{\sigma'}(\boldsymbol{v'},t) \\
&= \sum_{{\sigma'}} \frac{m_\sigma}{D} \int d^D\boldsymbol{v}\; \boldsymbol{v}^2\frac{\partial \,}{\partial \boldsymbol{v}}\cdot \int d^D\boldsymbol{v}'\; \left[\frac{m_{\sigma'} w_{\sigma'}}{k_B T_{\sigma, \parallel}}  \boldsymbol{Q}_{\sigma {\sigma'}}\cdot\boldsymbol{v} -\frac{m_{\sigma'} w_{\sigma}}{k_B T_{\sigma', \parallel}} \boldsymbol{Q'}_{\sigma {\sigma'}}\cdot\boldsymbol{v'} \right] f_\sigma(\boldsymbol{v},t)f_{\sigma'}(\boldsymbol{v'},t) \\
&= -\sum_{{\sigma'}} \frac{2m_\sigma}{D}  \int d^D\boldsymbol{v}\; \boldsymbol{v}\cdot \int d^D\boldsymbol{v}'\; \left[\frac{m_{\sigma'} w_{\sigma'}}{k_B T_{\sigma, \parallel}}  \boldsymbol{Q}_{\sigma {\sigma'}}\cdot\boldsymbol{v} -\frac{m_{\sigma'} w_{\sigma}}{k_B T_{\sigma', \parallel}} \boldsymbol{Q'}_{\sigma {\sigma'}}\cdot\boldsymbol{v'} \right] f_\sigma(\boldsymbol{v},t)f_{\sigma'}(\boldsymbol{v'},t)  \\
&= -\sum_{{\sigma'}} \frac{2m_\sigma m_{\sigma'}}{D}  \iint d^D\boldsymbol{v} d^D\boldsymbol{v}'\; \left[\frac{w_{\sigma'}}{k_B T_{\sigma, \parallel}}  \boldsymbol{v}\cdot \boldsymbol{Q}_{\sigma {\sigma'}}\cdot\boldsymbol{v} -\frac{w_{\sigma}}{k_B T_{\sigma', \parallel}} \boldsymbol{v}\cdot \boldsymbol{Q'}_{\sigma {\sigma'}}\cdot\boldsymbol{v'} \right] f_\sigma(\boldsymbol{v},t)f_{\sigma'}(\boldsymbol{v'},t)
\end{align}
\end{widetext}
where we used the assumption of Maxwellian distributions to replace the terms $\partial_{\boldsymbol{v}}f_{\sigma}$ in the second line, used integration by parts for the integral over $\boldsymbol{v}$ in the third line, and slightly rearranged terms in the fourth line. Using the explicit expressions of $\boldsymbol{Q}_{\sigma {\sigma'}}$ and $\boldsymbol{Q'}_{\sigma {\sigma'}}$ from \cref{eq:BGLoperator1_anyD,eq:BGLoperator2_anyD} while neglecting time aliases, this can be written as:
\begin{widetext}
\begin{align}
\frac{d\,k_BT_{\sigma, \parallel}}{dt} =& \sum_{{\sigma'}} \frac{2 \pi m_\sigma m_{\sigma'} \omega_{p,\sigma}^2 \omega_{p,{\sigma'}}^2}{D n_\sigma} \int_g\frac{d^D\boldsymbol{k}}{(2\pi)^D}\sum_{\boldsymbol{p} \boldsymbol{p'}}\;\iint d^D\boldsymbol{v}\,d^D\boldsymbol{v}'\frac{S_m^2(\boldsymbol{k}_{\boldsymbol{p}})S_m^2(\boldsymbol{k}_{\boldsymbol{p'}})}{|\epsilon(\boldsymbol{k}, \boldsymbol{k}_{\boldsymbol{p}}\cdot\boldsymbol{v})|^2 }
\times\nonumber\\
&\qquad \frac{(\boldsymbol{k}_{\boldsymbol{p}}\cdot\boldsymbol{v})}{K^4(\boldsymbol{k})} \left[ \frac{w_{\sigma'}(\boldsymbol{k}_{\boldsymbol{p}}\cdot\boldsymbol{v})}{k_B T_{\sigma, \parallel}} - \frac{w_{\sigma}(\boldsymbol{k}_{\boldsymbol{p'}}\cdot\boldsymbol{v}')}{k_B T_{\sigma', \parallel}} \right] \delta(\boldsymbol{k}_{\boldsymbol{p}}\cdot\boldsymbol{v}-\boldsymbol{k}_{\boldsymbol{p'}}\cdot\boldsymbol{v'}) f_\sigma(\boldsymbol{v},t)f_{\sigma'}(\boldsymbol{v'},t)\\
=& \sum_{{\sigma'}} \frac{2 \pi m_\sigma m_{\sigma'} \omega_{p,\sigma}^2 \omega_{p,{\sigma'}}^2}{D n_\sigma} \int_g\frac{d^D\boldsymbol{k}}{(2\pi)^D}\sum_{\boldsymbol{p} \boldsymbol{p'}}\;\iint d^D\boldsymbol{v}\,d^D\boldsymbol{v}'\frac{S_m^2(\boldsymbol{k}_{\boldsymbol{p}})S_m^2(\boldsymbol{k}_{\boldsymbol{p'}})}{|\epsilon(\boldsymbol{k}, \boldsymbol{k}_{\boldsymbol{p}}\cdot\boldsymbol{v})|^2}
\times\nonumber\\
&\qquad \frac{(\boldsymbol{k}_{\boldsymbol{p}}\cdot\boldsymbol{v})^2}{K^4(\boldsymbol{k})} \left[ \frac{w_{\sigma'}}{k_B T_{\sigma, \parallel}} - \frac{w_{\sigma}}{k_B T_{\sigma', \parallel}} \right] \delta(\boldsymbol{k}_{\boldsymbol{p}}\cdot\boldsymbol{v}-\boldsymbol{k}_{\boldsymbol{p'}}\cdot\boldsymbol{v'}) f_\sigma(\boldsymbol{v},t)f_{\sigma'}(\boldsymbol{v'},t)\\
=& \sum_{{\sigma'}}  \frac{2\omega_{p,\sigma}^2 \omega_{p,{\sigma'}}^2 \Delta x^{3-D}}{D(2\pi)^{3/2} n_\sigma} \frac{ [w_{\sigma'} k_B T_{\sigma', \parallel} - w_\sigma k_B T_{\sigma, \parallel}]\log \Lambda_{PIC, \sigma \sigma'} }{\left[ \, k_B T_{\sigma, \parallel}/m_\sigma + k_B T_{\sigma', \parallel}/m_{\sigma'}\,\right]^{3/2}}
\end{align}
\end{widetext}
where, in the second line, we used the property of $\delta(\boldsymbol{k}_{\boldsymbol{p}}\cdot\boldsymbol{v}-\boldsymbol{k}_{\boldsymbol{p'}}\cdot\boldsymbol{v'})$ to replace the term $(\boldsymbol{k}_{\boldsymbol{p'}}\cdot\boldsymbol{v'})$ by $(\boldsymbol{k}_{\boldsymbol{p}}\cdot\boldsymbol{v})$ in the integrand. In the third line, we rewrote the equation in a form similar to that of Spitzer theory, by defining $\log \Lambda_{PIC, \sigma \sigma'}$:
\begin{widetext}   
\begin{align}
\log \Lambda_{PIC, \sigma \sigma'}
\equiv& \frac{\left[ \, k_B T_{\sigma, \parallel}/m_\sigma + k_B T_{\sigma', \parallel}/m_{\sigma'}\,\right]^{3/2}}{k_B^2 T_{\sigma, \parallel} T_{\sigma', \parallel}/m_\sigma m_{\sigma'}}\int_g\frac{d^D\boldsymbol{k} \; \Delta x^{D-3}}{2^{D-3/2}\pi^{D-5/2}}\times \nonumber\\
&\quad \sum_{\boldsymbol{p} \boldsymbol{p'}}\;\iint d^D\boldsymbol{v}\,d^D\boldsymbol{v}'\frac{S_m^2(\boldsymbol{k}_{\boldsymbol{p}})S_m^2(\boldsymbol{k}_{\boldsymbol{p'}})}{|\epsilon(\boldsymbol{k}, \boldsymbol{k}_{\boldsymbol{p}}\cdot\boldsymbol{v})|^2}\frac{(\boldsymbol{k}_{\boldsymbol{p}}\cdot\boldsymbol{v})^2}{K^4(\boldsymbol{k})}
\delta(\boldsymbol{k}_{\boldsymbol{p}}\cdot\boldsymbol{v}-\boldsymbol{k}_{\boldsymbol{p'}}\cdot\boldsymbol{v'})f_\sigma(\boldsymbol{v},t)f_{\sigma'}(\boldsymbol{v'},t)
\end{align}
\end{widetext}

In the integral over $\boldsymbol{v}$, $\boldsymbol{v'}$ we now decompose the components parallel and orthogonal to $\boldsymbol{k}_{\boldsymbol{p}}$ and $\boldsymbol{k}_{\boldsymbol{p'}}$ respectively: $v_\parallel$, $\boldsymbol{v}_\perp$ and $v'_\parallel$, $\boldsymbol{v'}_\perp$. ($\boldsymbol{v}_\perp$ and $\boldsymbol{v}'_\perp$ are of dimension $(D-1)$ and do not exist for $D=1$, in which case the formulas below remain valid by simply skipping the integral over $\boldsymbol{v}_\perp$ and $\boldsymbol{v}'_\perp$) With these notations, $\boldsymbol{k}_{\boldsymbol{p}}\cdot\boldsymbol{v} = k_{\boldsymbol{p}}v_\parallel$ and $\boldsymbol{k}_{\boldsymbol{p'}}\cdot\boldsymbol{v'} = k_{\boldsymbol{p'}}v'_\parallel$.

\begin{widetext}    
\begin{align}
\log &\Lambda_{PIC, \sigma \sigma'} = \frac{\left[ \, k_B T_{\sigma, \parallel}/m_\sigma + k_B T_{\sigma', \parallel}/m_{\sigma'}\,\right]^{3/2}}{k_B^2 T_{\sigma, \parallel} T_{\sigma', \parallel}/m_\sigma m_{\sigma'}}\int_g\frac{d^D\boldsymbol{k}\;\Delta x^{D-3}}{2^{D-3/2}\pi^{D-5/2}}\times\nonumber\\
&\qquad\sum_{\boldsymbol{p} \boldsymbol{p'}}\;\iint d^{D-1}\boldsymbol{v}_\perp dv_\parallel\,d^{D-1}\boldsymbol{v}'_\perp dv'_\parallel \frac{S_m^2(\boldsymbol{k}_{\boldsymbol{p}})S_m^2(\boldsymbol{k}_{\boldsymbol{p'}})}{|\epsilon(\boldsymbol{k}, k_{\boldsymbol{p}} v_\parallel)|^2}\frac{k_{\boldsymbol{p}}^2 v_\parallel^2}{K^4(\boldsymbol{k})}
\delta(k_{\boldsymbol{p}}v_\parallel-k_{\boldsymbol{p'}}v'_\parallel)f_\sigma(\boldsymbol{v},t)f_{\sigma'}(\boldsymbol{v'},t)\\
=& \frac{\left[ \, k_B T_{\sigma, \parallel}/m_\sigma + k_B T_{\sigma', \parallel}/m_{\sigma'}\,\right]^{3/2}}{k_B^2 T_{\sigma, \parallel} T_{\sigma', \parallel}/m_\sigma m_{\sigma'}}\int_g\frac{d^D\boldsymbol{k}\;\Delta x^{D-3}}{2^{D-3/2}\pi^{D-5/2}} \times\nonumber \\
&\qquad\sum_{\boldsymbol{p} \boldsymbol{p'}}\;\iint dv_\parallel dv'_\parallel \frac{S_m^2(\boldsymbol{k}_{\boldsymbol{p}})S_m^2(\boldsymbol{k}_{\boldsymbol{p'}})}{|\epsilon(\boldsymbol{k}, k_{\boldsymbol{p}} v_\parallel)|^2}\frac{k_{\boldsymbol{p}}^2 v_\parallel^2}{K^4(\boldsymbol{k})}
\delta(k_{\boldsymbol{p}}v_\parallel-k_{\boldsymbol{p'}}v'_\parallel)\frac{e^{-m_\sigma v_\parallel^2/2k_BT_{\sigma, \parallel} - m_{\sigma'}v'^2_\parallel/2k_BT_{\sigma', \parallel}}}{(2\pi k_B T_{\sigma, \parallel}/m_\sigma )^{1/2}(2\pi k_B T_{\sigma', \parallel}/m_{\sigma'} )^{1/2}}\\
=& \left[ \frac{m_\sigma}{k_B T_{\sigma, \parallel}} + \frac{m_{\sigma'}}{k_B T_{\sigma', \parallel}}\right]^{3/2}\int_g\frac{d^D\boldsymbol{k}\;\Delta x^{D-3}}{2^{D-1/2}\pi^{D-3/2}}\times \nonumber \\
&\qquad \sum_{\boldsymbol{p} \boldsymbol{p'}}\;\int dv_\parallel \frac{S_m^2(\boldsymbol{k}_{\boldsymbol{p}})S_m^2(\boldsymbol{k}_{\boldsymbol{p'}})}{|\epsilon(\boldsymbol{k}, k_{\boldsymbol{p}} v_\parallel)|^2}\frac{k_{\boldsymbol{p}}^2v_\parallel^2}{K^4(\boldsymbol{k}) k_{\boldsymbol{p'}}}e^{-\left[\frac{m_\sigma}{2k_BT_{\sigma, \parallel} k_{\boldsymbol{p}}^2} + \frac{m_{\sigma'} }{2k_BT_{\sigma', \parallel}k_{\boldsymbol{p'}}^2}\right]k_{\boldsymbol{p}}^2v^2_\parallel}\\
=& \int_g\frac{d^D\boldsymbol{k}\;\Delta x^{D-3}}{2^{D-2}\pi^{D-3/2}}\sum_{\boldsymbol{p} \boldsymbol{p'}} \frac{S_m^2(\boldsymbol{k}_{\boldsymbol{p}})S_m^2(\boldsymbol{k}_{\boldsymbol{p'}})}{K^4(\boldsymbol{k}) k_{\boldsymbol{p}}k_{\boldsymbol{p'}}}\frac{\left[ \frac{m_\sigma}{k_B T_{\sigma, \parallel}} + \frac{m_{\sigma'}}{k_B T_{\sigma', \parallel}}\right]^{3/2}}{\left[ \frac{m_\sigma}{k_B T_{\sigma, \parallel} k_{\boldsymbol{p}}^2} + \frac{m_{\sigma'}}{k_B T_{\sigma', \parallel}k_{\boldsymbol{p'}}^2}\right]^{3/2}}\times \\ \nonumber
&\qquad\int d\alpha\frac{\alpha^2e^{-\alpha^2}}{\left|\epsilon\left(\boldsymbol{k}, \alpha\left(\frac{m_\sigma}{2k_B T_{\sigma, \parallel} k_{\boldsymbol{p}}^2} + \frac{m_{\sigma'}}{2k_B T_{\sigma', \parallel}k_{\boldsymbol{p'}}^2}\right)^{-1/2}\right) \right|^2}
\end{align}
\end{widetext}
where we used the assumption of Maxwellian distribution (\cref{eq:Maxwellian_anyD}) and carried out the integrals over $\boldsymbol{v}_\perp$, $\boldsymbol{v}_\perp'$ in the second line.  We carried out the integral over $v_\parallel'$ in the third line, by using the property of the $\delta$ function. We performed the change of variable $k_{\boldsymbol{p}}v_\parallel = \alpha \left(\frac{m_\sigma}{2k_B T_{\sigma, \parallel} k_{\boldsymbol{p}}^2} + \frac{m_{\sigma'}}{2k_B T_{\sigma', \parallel}k_{\boldsymbol{p'}}^2}\right)^{-1/2}$ in the integral on the fourth line.

The expression of $\epsilon(\boldsymbol{k}, \omega)$ from \cref{eq:dielectric_anyD} can be further simplified by neglecting the time aliases: this can be seen by writing the $\cot$ term in \cref{eq:dielectric_anyD} as a sum over time aliases: $\Delta t/2 \times \cot[  (\omega -\boldsymbol{k}_{\boldsymbol{p''}}\cdot\boldsymbol{v}'')\Delta t/2 ] = \sum_{q=-\infty}^{\infty} 1/[ (\omega -\boldsymbol{k}_{\boldsymbol{p''}}\cdot\boldsymbol{v}'')-2\pi q/\Delta t ]$ \cite{Abramowitz} and retaining only the term for $q=0$, i.e. $\Delta t/2 \times \cot[  (\omega -\boldsymbol{k}_{\boldsymbol{p''}}\cdot\boldsymbol{v}'')\Delta t/2 ] \approx 1/ (\omega -\boldsymbol{k}_{\boldsymbol{p''}}\cdot\boldsymbol{v}'')$
\begin{widetext}
\begin{align}
\epsilon(\boldsymbol{k}, \omega) &= 1 + \sum_{\sigma''} \frac{\omega_{p,\sigma''}^2}{K^2(\boldsymbol{k})}\sum_{\boldsymbol{p''}} S_m^2(\boldsymbol{k}_{\boldsymbol{p''}})\int d^D\boldsymbol{v}'' \;\boldsymbol{k}_{\boldsymbol{p''}}\cdot\frac{\partial f_{\sigma''}(\boldsymbol{v}'',t)}{\partial \boldsymbol{v}''}\frac{1}{\omega - \boldsymbol{k}_{\boldsymbol{p''}}\cdot \boldsymbol{v}''} \\
&= 1 - \sum_{\sigma''} \frac{m_{\sigma''}\omega_{p,\sigma''}^2}{K^2(\boldsymbol{k}) k_B T_{\sigma'', \parallel}}\sum_{\boldsymbol{p''}} S_m^2(\boldsymbol{k}_{\boldsymbol{p''}})\int d^D\boldsymbol{v}'' \;\frac{\boldsymbol{k}_{\boldsymbol{p''}}\cdot \boldsymbol{v}''}{(\omega - \boldsymbol{k}_{\boldsymbol{p''}}\cdot \boldsymbol{v}'')} \frac{e^{ -\frac{m_{\sigma''}\boldsymbol{v}''^2}{2k_B T_{\sigma'', \parallel}}} }{(2\pi k_B T_{\sigma'', \parallel}/m_{\sigma''})^{D/2} }\\
&= 1 - \sum_{\sigma''} \frac{m_{\sigma''}\omega_{p,\sigma''}^2}{K^2(\boldsymbol{k}) k_B T_{\sigma'', \parallel}}\sum_{\boldsymbol{p''}} S_m^2(\boldsymbol{k}_{\boldsymbol{p''}})\int d v_\parallel'' \;\frac{k_{\boldsymbol{p''}} v_\parallel''}{(\omega - k_{\boldsymbol{p''}}v_\parallel'')} \frac{e^{ -\frac{m_{\sigma''}v_\parallel''^2}{2k_B T_{\sigma'', \parallel}}}}{(2\pi k_B T_{\sigma'', \parallel}/m_{\sigma''})^{1/2} } \\
&= 1 - \sum_{\sigma''} \frac{m_{\sigma''}\omega_{p,\sigma''}^2}{K^2(\boldsymbol{k}) k_B T_{\sigma'', \parallel}}\sum_{\boldsymbol{p''}} S_m^2(\boldsymbol{k}_{\boldsymbol{p''}})\int \frac{d u}{\pi^{1/2} } \;\frac{u\, e^{-u^2}}{[\,\omega (m_{\sigma''}/2 k_B T_{\sigma'', \parallel})^{1/2}/k_{\boldsymbol{p''}}  - u\,]} \\
&= 1 - \sum_{\sigma''} \frac{m_{\sigma''}\omega_{p,\sigma''}^2}{2K^2(\boldsymbol{k}) k_B T_{\sigma'', \parallel}}\sum_{\boldsymbol{p''}} S_m^2(\boldsymbol{k}_{\boldsymbol{p''}}) Z'\left( \omega \left(\frac{m_{\sigma''}}{2 k_B T_{\sigma'', \parallel}k_{\boldsymbol{p''}}^2}\right)^{1/2} \right)
\end{align}
\end{widetext}
where we used the assumption that $f_{\sigma''}$ is a Maxwellian (\cref{eq:Maxwellian_anyD}), in the second line, and where we again decomposed the integral over $\boldsymbol{v}''$ between the components $v''_\parallel$ and $\boldsymbol{v}''_\perp$, which are parallel and orthogonal to $\boldsymbol{k}_{\boldsymbol{p''}}$ respectively. Similarly to \cite{Okuda,Langdon}, we find that $\epsilon(\boldsymbol{k}, \omega)$ can be expressed with the plasma dispersion function \cite{FriedConte1961}
\begin{equation}
Z'(\zeta) \equiv \frac{1}{\sqrt{\pi}}\int_{-\infty}^{\infty}du \frac{e^{-u^2}}{(\zeta -u)^2} = \frac{1}{\sqrt{\pi}}\int_{-\infty}^{\infty}du \frac{2u\, e^{-u^2}}{(\zeta-u)}
\label{eq:plasma_dispersion}
\end{equation}

In summary, the equations governing the temperature evolution in any dimension $D$, including the influence of spatial aliases, are:
\begin{widetext}
\begin{align}
\frac{d\,T_{\sigma, \parallel}}{dt} &= \sum_{{\sigma'}}  \frac{2\omega_{p,\sigma}^2 \omega_{p,{\sigma'}}^2 \Delta x^{3-D}}{D(2\pi)^{3/2} n_\sigma} \frac{ [w_{\sigma'} T_{\sigma', \parallel} - w_\sigma T_{\sigma, \parallel}]\log \Lambda_{PIC, \sigma \sigma'} }{\left[ \, k_B T_{\sigma, \parallel}/m_\sigma + k_B T_{\sigma', \parallel}/m_{\sigma'}\,\right]^{3/2}} \label{eq:evolutionT_anyD_with_aliases} \\
\log \Lambda_{PIC, \sigma \sigma'} &= \int_g\frac{d^D\boldsymbol{k}\;\Delta x^{D-3}}{2^{D-2}\pi^{D-3/2}}\sum_{\boldsymbol{p} \boldsymbol{p'}} \frac{S_m^2(\boldsymbol{k}_{\boldsymbol{p}})S_m^2(\boldsymbol{k}_{\boldsymbol{p'}})}{K^4(\boldsymbol{k}) k_{\boldsymbol{p}}k_{\boldsymbol{p'}}}\frac{\left[ \frac{m_\sigma}{k_B T_{\sigma, \parallel}} + \frac{m_{\sigma'}}{k_B T_{\sigma', \parallel}}\right]^{3/2}}{\left[ \frac{m_\sigma}{k_B T_{\sigma, \parallel} k_{\boldsymbol{p}}^2} + \frac{m_{\sigma'}}{k_B T_{\sigma', \parallel}k_{\boldsymbol{p'}}^2}\right]^{3/2}} \times \nonumber \\
&\qquad \int_{-\infty}^\infty \!\!\!\!\!d\alpha\frac{\alpha^2e^{-\alpha^2}}{\left|\epsilon\left(\boldsymbol{k}, \alpha \left(\frac{m_\sigma}{2k_B T_{\sigma, \parallel} k_{\boldsymbol{p}}^2} + \frac{m_{\sigma'}}{2k_B T_{\sigma', \parallel}k_{\boldsymbol{p'}}^2}\right)^{-1/2}\right) \right|^2} 
\label{eq:logPIC_anyD_with_aliases} \\
\epsilon(\boldsymbol{k}, \omega) &= 1 - \sum_{\sigma''} \frac{m_{\sigma''}\omega_{p,\sigma''}^2}{2K^2(\boldsymbol{k}) k_B T_{\sigma'', \parallel}}\sum_{\boldsymbol{p''}} S_m^2(\boldsymbol{k}_{\boldsymbol{p''}}) Z'\left( \omega \left(\frac{m_{\sigma''}}{2 k_B T_{\sigma'', \parallel} k_{\boldsymbol{p''}}^2}\right)^{1/2}  \right)
\label{eq:dielectricM_anyD_with_aliases} 
\end{align}
\end{widetext}

The influence of spatial aliases is discussed further in \cref{app:aliases}; in the main text and in all other appendices, they are neglected. This amounts to retaining only the terms $\boldsymbol{p} = 0$, $\boldsymbol{p'} = 0$ and $\boldsymbol{p''} = 0$ in the sums over $\boldsymbol{p}$, $\boldsymbol{p'}$ and $\boldsymbol{p''}$ in \cref{eq:logPIC_anyD_with_aliases,eq:dielectricM_anyD_with_aliases}, and yields \cref{eq:evolutionT_anyD,eq:logPIC_anyD,eq:dielectricM_anyD}. Specializing further to $D=3$, in which case the temperature along the grid axes $T_{\sigma,\parallel}$ is simply the temperature $T_\sigma$, this system reduces to \cref{eq:evolutionT,eq:logPIC,eq:dielectricM}.

\section{Approximate expression of $\log \Lambda_{PIC}$ for an electron-ion plasma with similar temperatures $T_{e,\parallel} \sim T_{i,\parallel}$}
\label{app:approximate}

\begin{figure*}
\includegraphics[width=\textwidth]{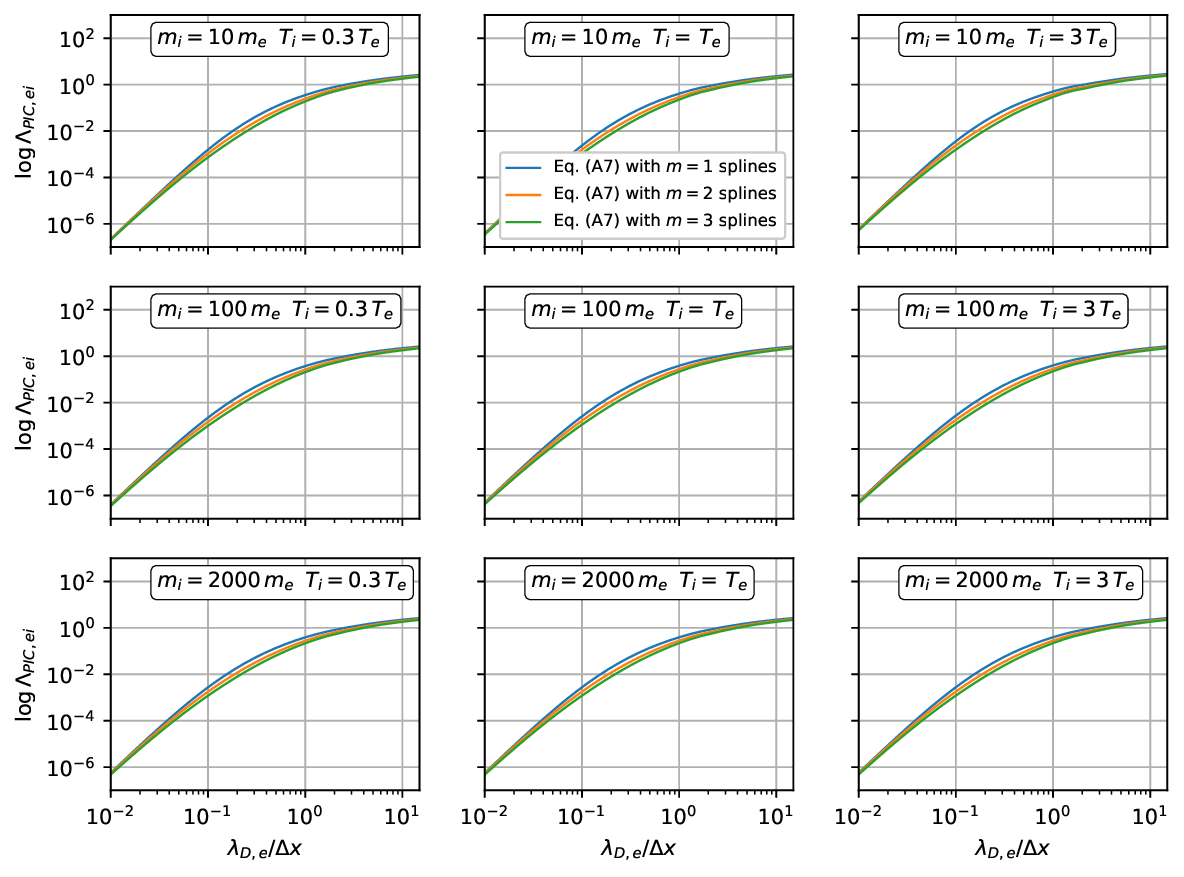}
\caption{Plots of $\log\Lambda_{PIC, e i}$ (as given by \cref{eq:logPIC_anyD} for $D=3$) as a function of $\lambda_{D,e}/\Delta x$, for different values of the ion mass $m_i$ (different rows) and of the ion temperature $T_i$ (different columns), as well as for different spline orders (different colors). The curves show that  $\Lambda_{PIC, e i}$ is essentially independent of $m_i$ and $T_i$ in this range of parameters.\label{fig:logPIC_massT}}
\end{figure*}

In order to find an approximated expression of $\log \Lambda_{PIC, \sigma \sigma'}$ from \cref{eq:logPIC_anyD}, we consider a two-species plasma made of electrons and ions ($m_e \ll m_i$) with similar temperature $T_{e,\parallel} \sim T_{i,\parallel}$. (See \cref{app:2d1d} for a definition of $T_{\sigma,\parallel}$.) In this case, when evaluating $\epsilon\left(\boldsymbol{k}, \alpha k /\sqrt{\frac{m_e}{2k_B T_{e,\parallel}} + \frac{m_{i}}{2k_B T_{i,\parallel}} } \right)$ in \cref{eq:logPIC_anyD}, the argument of the $Z'$ function in \cref{eq:dielectricM_anyD} can be approximated by assuming $m_i/T_{i,\parallel} \gg m_e/T_{e,\parallel}$:
\begin{align} 
Z'\left(\alpha \frac{\left(\frac{m_{i}}{ k_B T_{i,\parallel}}\right)^{1/2}}{\left(\frac{m_{e}}{ k_B T_{e,\parallel}}+\frac{m_{i}}{ k_B T_{i,\parallel}}\right)^{1/2}} \right) &\approx Z'(\alpha)
\label{eq:approx_ion}
\\
Z'\left(\alpha \frac{\left(\frac{m_{e}}{ k_B T_{e,\parallel}}\right)^{1/2}}{\left(\frac{m_{e}}{ k_B T_{e,\parallel}}+\frac{m_{i}}{ k_B T_{i,\parallel}}\right)^{1/2}} \right) &\approx Z'(0) = -2
\label{eq:approx_electron}
\end{align}
The approximations \cref{eq:approx_ion,eq:approx_electron} are only valid for $T_{i,\parallel} \sim T_{e,\parallel}$. This is because, if $T_{i,\parallel} \gg T_{e,\parallel}$, then the assumption $m_i/T_{i,\parallel} \gg m_e/T_{e,\parallel}$ might not be valid anymore. On the other hand, if $T_{i,\parallel} \ll T_{e,\parallel}$, further analysis (not detailed here) shows that the integral over $\alpha$ in \cref{eq:logPIC_anyD} has significant contributions from large values of $\alpha$, where \cref{eq:approx_electron} is not valid anymore.

Let us the short-hand notation
\begin{equation}
\tilde{\epsilon}(\boldsymbol{k}, \alpha) = \epsilon\left(\boldsymbol{k}, \frac{\alpha k}{\sqrt{ \frac{m_e}{2k_B T_{e,\parallel}} + \frac{m_{i}}{2k_B T_{i,\parallel}} }} \right) 
\end{equation}
Using \cref{eq:approx_ion,eq:approx_electron}, the expression of $\tilde{\epsilon}(\boldsymbol{k}, \alpha)$ can be simplified to
\begin{equation}
\tilde{\epsilon}(\boldsymbol{k}, \alpha) = 1 + \frac{S_m^2(\boldsymbol{k})}{K^2(\boldsymbol{k}) \lambda_{D,e}^2} -  \frac{S_m^2(\boldsymbol{k})}{2 K^2(\boldsymbol{k}) \lambda_{D,i}^2} Z' (\alpha)
\label{eq:approx_epsilonM}
\end{equation}

In this case, we can then use techniques similar to \cite{Birdsall} (section 12-3-b) to compute the integral over $\alpha$ in the expression \cref{eq:logPIC_anyD} of $\log \Lambda_{PIC,ei}$. More specifically, this integral over $\alpha$ can be rewritten as the imaginary part $\Im$ of the integral of an analytical function, for which we can then use contour integrals to obtain the final result. In order to do so, we first use the properties of the plasma dispersion function $Z'(\alpha) = -2 [ 1 + \alpha Z(\alpha) ] $ and (for $\alpha$ real) $\Im[\,Z(\alpha)\,] = \sqrt{\pi}e^{-\alpha^2}$ \cite{FriedConte1961} to express $\Im[ \, Z'(\alpha) \, ]$:
\begin{equation}
\Im[ \, Z'(\alpha) \, ] = -2\sqrt{\pi} \, \alpha e^{-\alpha^2} 
\end{equation}
and therefore from \cref{eq:approx_epsilonM} 
\begin{equation}
\Im\left[ \tilde{\epsilon}(\boldsymbol{k}, \alpha) \right] = \frac{\sqrt{\pi}S_m^2(\boldsymbol{k})}{K^2(\boldsymbol{k}) \lambda_{D,i}^2}  \alpha e^{-\alpha^2} 
\end{equation}
Thus, \cref{eq:logPIC_anyD} can be written as:
\begin{widetext}
\begin{align}
\log \Lambda_{PIC, ei} &= \int_g\frac{d^D\boldsymbol{k}\;\Delta x^{D-3}}{2^{D-2}\pi^{D-1}}\frac{S_m^2(\boldsymbol{k}) k\lambda_{D,i}^2}{K^2(\boldsymbol{k})} \int_{-\infty}^\infty \!\!\!\!\!d\alpha\;\alpha\frac{\Im [\, \tilde{\epsilon}(\boldsymbol{k}, \alpha) \,]}{\left|\tilde{\epsilon}(\boldsymbol{k}, \alpha)\right|^2} \\
&= -\int_g\frac{d^D\boldsymbol{k}\;\Delta x^{D-3}}{2^{D-2}\pi^{D-1}}\frac{S_m^2(\boldsymbol{k}) k\lambda_{D,i}^2}{K^2(\boldsymbol{k})} \Im\left[ \int_{-\infty}^\infty \!\!\!\!\!d\alpha \; \alpha\frac{1}{\tilde{\epsilon}(\boldsymbol{k}, \alpha)} \right] \\
&= -\int_g\frac{d^D\boldsymbol{k}\;\Delta x^{D-3}}{2^{D-2}\pi^{D-1}}\frac{S_m^2(\boldsymbol{k}) k\lambda_{D,i}^2}{K^2(\boldsymbol{k})} \Im\left[ \int_{-\infty}^\infty \!\!\!\!\!d\alpha \; \alpha\left( \frac{1}{\tilde{\epsilon}(\boldsymbol{k}, \alpha)} - \frac{1}{1 + \frac{S_m^2(\boldsymbol{k})}{K^2(\boldsymbol{k}) \lambda_{D,e}^2}} \right)\right]  \label{eq:int_analytical_function}
\end{align}
\end{widetext}
where, in the second line, we used the relation $\Im[\,1/z \,] = -\Im[z]/|z|^2$, which is valid for any complex number $z$, and where, in the third line, we deliberately added a purely real function in the integrand, which is valid since it does not contribute to the imaginary part $\Im$. In \cref{eq:int_analytical_function}, the integral over $\alpha$ is the integral of an analytical function:
\begin{equation}
f(\alpha) = \alpha\left( \frac{1}{\tilde{\epsilon}(\boldsymbol{k}, \alpha)} - \frac{1}{1 + \frac{S_m^2(\boldsymbol{k})}{K^2(\boldsymbol{k}) \lambda_{D,e}^2}} \right) 
\label{eq:analytical}
\end{equation}
Assuming that this analytical function has no pole in the upper half of the complex plane, its integral over a closed contour in that upper half is 0:
\begin{equation}
\int_{-\infty}^\infty d\alpha\,f(\alpha) + \int_{\mathcal{C}^+} d\alpha \,f(\alpha) = 0
\end{equation}
where $\mathcal{C}^+$ is a semi-circle at large $\alpha$ in the upper half of the complex plane. Expressing the integral over $\mathcal{C}^+$ requires knowing the asymptotic behavior of $f(\alpha)$ for $|\alpha| \rightarrow \infty$. Using the properties of the plasma dispersion function $Z'(\alpha) = -2 [ 1 + \alpha Z(\alpha) ] $ and $Z(\alpha) \approx -1/\alpha-1/(2\alpha^3)$ for $|\alpha| \rightarrow \infty$ \cite{FriedConte1961}, we have:
\begin{equation}
Z'(\alpha) \sim \frac{1}{\alpha^2} \quad \mathrm{for} \; |\alpha| \rightarrow \infty
\end{equation} 
Combining this relation with \cref{eq:analytical} and \cref{eq:approx_epsilonM} gives
\begin{equation}
f(\alpha) \sim \frac{\frac{S_m^2(\boldsymbol{k})}{2 K^2(\boldsymbol{k}) \lambda_{D,i}^2}}{\left(1 + \frac{S_m^2(\boldsymbol{k})}{K^2(\boldsymbol{k}) \lambda_{D,e}^2}\right)^2}\frac{1}{\alpha}  \quad \mathrm{for} \; |\alpha| \rightarrow \infty
\end{equation}
and thus
\begin{align}
\int_{-\infty}^\infty &d\alpha\,f(\alpha) = -\int_{\mathcal{C}^+} d\alpha \,f(\alpha)\\
&= -\frac{\frac{S_m^2(\boldsymbol{k})}{2 K^2(\boldsymbol{k}) \lambda_{D,i}^2}}{\left(1 + \frac{S_m^2(\boldsymbol{k})}{K^2(\boldsymbol{k}) \lambda_{D,e}^2}\right)^2} \int_{\mathcal{C}^+} \frac{d\alpha}{\alpha}\\
&= -\frac{\frac{S_m^2(\boldsymbol{k})}{2 K^2(\boldsymbol{k}) \lambda_{D,i}^2}}{\left(1 + \frac{S_m^2(\boldsymbol{k})}{K^2(\boldsymbol{k}) \lambda_{D,e}^2}\right)^2} i\pi
\end{align}
Finally, inserting this expression into \cref{eq:int_analytical_function} yields
\begin{equation}
\log \Lambda_{PIC, e i} = \int_g\frac{d^D\boldsymbol{k}\;\Delta x^{D-3}}{2^{D-1}\pi^{D-2}}\frac{\frac{S_m^4(\boldsymbol{k}) k}{K^4(\boldsymbol{k})}}{\left( 1 + \frac{S_m^2(\boldsymbol{k})}{K^2(\boldsymbol{k}) \lambda_{D,e}^2}\right)^2}
\label{eq:logPIC_two_species}
\end{equation}

Surprisingly, despite the fact that \cref{eq:logPIC_anyD,eq:approx_epsilonM} depend on the ion Debye length $\lambda_{D,i}$, this dependency cancels out in the above final expression for $\log \Lambda_{PIC, e i}$. Instead, in \cref{eq:logPIC_two_species}, $\log \Lambda_{PIC, e i}$ depends only on the electron Debye length $\lambda_{D,e}$. If we further assume, for the remainder of this appendix, that the cell sizes are equal in all directions ($\Delta x = \Delta y = \Delta z$), then $\Delta x$ and $\lambda_{D,e}$ are the only lengths entering $\log \Lambda_{PIC, e i}$. Since $\log \Lambda_{PIC, e i}$ is dimensionless, it can therefore depend only on the ratio $\lambda_{D,e}/\Delta x$. This is confirmed by \cref{fig:logPIC_massT}, in which \cref{eq:logPIC_anyD} was evaluated numerically for different values of $m_i$ and $T_i$, and which shows that $\log \Lambda_{PIC, ei}$ is independent of these values and only depends on $\lambda_{D,e}/\Delta x$.

The fact that $\log \Lambda_{PIC, e i}$ only depends on $\lambda_{D,e}/\Delta x$ motivates examining the limits $\lambda_{D,e}/\Delta x \ll 1$ and $\lambda_{D,e}/\Delta x \gg 1$, in which the expression simplifies further.

\subsection{Under-resolved Debye length ($\lambda_{D,e} / \Delta x \ll 1$)}

In this case, the term $S_m^2(\boldsymbol{k})/K^2(\boldsymbol{k})\lambda_{D,e}^2$ in \cref{eq:logPIC_two_species} (which is of order $\sim \Delta x^2/\lambda_{D,e}^2$) dominates the denominator over much of the integration interval for $\boldsymbol{k}$. We can thus consider $1 + S_m^2(\boldsymbol{k})/K^2(\boldsymbol{k})\lambda_{D,e}^2 \approx S_m^2(\boldsymbol{k})/K^2(\boldsymbol{k})\lambda_{D,e}^2$, which leads to the following simplified expression:
\begin{equation}
\log \Lambda_{PIC, ei} = \frac{\Delta x^{D-3}\lambda_{D,e}^4 }{2^{D-1}\pi^{D-2}}\int_g \, d^D\boldsymbol{k} \;k
\end{equation}
The integral $\int_g \, d^D\boldsymbol{k} \;k$ is the integral of $k \equiv \sqrt{\boldsymbol{k}^2}$ over the first Brillouin zone, which is a cube (for $D=3$), a square (for $D=2$) or a segment (for $D=1$) extending from $-\pi/\Delta x$ to $+\pi/\Delta x$ in each dimension. (Again, we assume that cell sizes are equal in all directions in the remainder of this appendix: $\Delta x = \Delta y = \Delta z$.) Thus, this integral can be expressed in terms of known $D$-dimensional ``box integrals'' $\mathcal{B}_D(1) \equiv \int_{[0,1]^D}d^D\boldsymbol{u}\,\sqrt{\boldsymbol{u}^2}$ \cite{Bailey}.
\[\int_g \, d^D\boldsymbol{k} \;k = 2^D\left( \frac{\pi}{\Delta x}\right)^{D+1} \mathcal{B}_D(1)\]
where, from \cite{Bailey}: 
\begin{align}
\mathcal{B}_1(1) &= \frac{1}{2} \label{eq:B1}\\
\mathcal{B}_2(1) &= \frac{\sqrt{2}}{3} + \frac{1}{3}\log(\sqrt{2}+1) \label{eq:B2}\\
\mathcal{B}_3(1) &= \frac{\sqrt{3}}{4} + \frac{1}{2}\log(2+\sqrt{3}) - \frac{\pi}{24} \label{eq:B3}
\end{align}
Therefore:
\begin{align}
\log \Lambda_{PIC, ei} &= 2\pi^3 \mathcal{B}_D(1) \left(\frac{\lambda_{D,e}}{\Delta x}\right)^4 \\
&\approx \left\{ \begin{array}{l l}
31 \times \left( \frac{\lambda_{D,e}}{\Delta x}\right)^4 & \mathrm{for}\,D=1\\ [8pt]
47 \times \left( \frac{\lambda_{D,e}}{\Delta x}\right)^4 & \mathrm{for}\,D=2\\[8pt]
60 \times \left( \frac{\lambda_{D,e}}{\Delta x}\right)^4 & \mathrm{for}\,D=3
\end{array}\right.
\end{align}

\subsection{Well-resolved Debye length ($\lambda_{D,e} / \Delta x \gg 1$)}

In this case, the main contribution to the integral is for $k \ll \pi/\Delta x$, where we can make the approximation $K(\boldsymbol{k}) \approx k, S_m(\boldsymbol{k}) \approx 1$. With these approximation, the integrand only depends on the magnitude $k$ of the $\boldsymbol{k}$ vector (and not on its direction) and thus $\int d^D\boldsymbol{k} = \frac{2\pi^{D/2}}{\Gamma(D/2)}\int k^{D-1} dk$.
\begin{align*}
\log\Lambda_{PIC} &= \frac{\Delta x^{D-3}}{2^{D-2}\pi^{D/2-2}\Gamma(D/2)} \int_0^{\sim \frac{1}{\Delta x}} \!\!\!\!\!\!\! \frac{ dk\; k^{D-4} }{\left( 1 + \frac{1}{(k\lambda_{D,e})^2} \right)^2} \\
 &= \frac{(\Delta x/\lambda_{D,e})^{D-3}}{2^{D-2}\pi^{D/2-2}\Gamma(D/2)} \int_0^{\sim \frac{\lambda_{D,e}}{\Delta x}} \!\!\!\!\!\!\! \frac{ du \; u^D }{\left( 1 + u^2\right)^2} 
\end{align*}
In the above, the upper bound $\sim 1/\Delta x$ on the integral reflects the fact that the approximations $K(\boldsymbol{k}) \approx k$ and $S_m(\boldsymbol{k}) \approx 1$ break down for $|\boldsymbol{k}| \sim 1/\Delta x$. Fortunately, as shown below, the final result is, to leading order, insensitive to the precise value of this cut-off.
The integral over $u$ can be calculated analytically for $D=1, 2, 3$:
\[
\log\Lambda_{PIC,ei} = \left\{ \begin{array}{l}
\pi \frac{\lambda_{D,e}^2}{\Delta x^2} \left[ -\frac{1}{1+u^2} \right]_0^{\sim \frac{\lambda_{D,e}}{\Delta x} }\\
\frac{\pi\lambda_{D,e}}{2\Delta x}\left[ \arctan(u)-\frac{u}{1+u^2}\right]_0^{\sim \frac{\lambda_{D,e}}{\Delta x} }\\
\frac{1}{2}\left[ \log(1+u^2) + \frac{1}{1+u^2} \right]_0^{\sim \frac{\lambda_{D,e}}{\Delta x} }
\end{array}\right.
\]
In the limit $\lambda_{D,e}/\Delta x \gg1$, the leading order term is:
\begin{equation}
\log\Lambda_{PIC,ei} \approx \left\{ \begin{array}{l l}
\pi \frac{\lambda_{D,e}^2}{\Delta x^2} & \mathrm{for}\,D=1\\ [8pt]
\frac{\pi^2}{4} \frac{\lambda_{D,e}}{\Delta x}  & \mathrm{for}\,D=2\\[8pt]
\log \left(\frac{\lambda_{D,e}}{\Delta x}\right) & \mathrm{for}\,D=3
\end{array}\right.
\end{equation}

\section{Influence of spatial aliases}
\label{app:aliases}

In the main text, \cref{fig:shape_factor} compared PIC simulations with the predictions of \cref{eq:evolutionT,eq:logPIC,eq:dielectricM}, where spatial aliases have been neglected, and found good agreement except for splines of order $m=1$. In this appendix, we revisit this comparison by using \cref{eq:evolutionT_anyD_with_aliases,eq:logPIC_anyD_with_aliases,eq:dielectricM_anyD_with_aliases} (which do take into account spatial aliases) instead of \cref{eq:evolutionT,eq:logPIC,eq:dielectricM}.

More specifically, \cref{eq:logPIC_anyD_with_aliases,eq:dielectricM_anyD_with_aliases} involve sums over the spatial aliases $\boldsymbol{p}$, $\boldsymbol{p'}$ and $\boldsymbol{p''}$, which are $D$-dimensional vectors with integer components. These sums contain infinitely many terms and must therefore be truncated when the expressions are evaluated numerically, e.g. when integrating the temperature evolution \cref{eq:evolutionT_anyD_with_aliases}. The truncation is justified by the rapid decay of the terms in \cref{eq:logPIC_anyD_with_aliases} at large $|\boldsymbol{p}|$ and $|\boldsymbol{p'}|$, which originates from the factor $S_m^2(\boldsymbol{k}_{\boldsymbol{p}})S_m^2(\boldsymbol{k}_{\boldsymbol{p'}})/k_{\boldsymbol{p}}k_{\boldsymbol{p'}}$. Recall from \cref{sec:evolution-distribution} that
$S_m(\boldsymbol{k}) = [\,\mathrm{sinc}(k_x\Delta x/2)\,\mathrm{sinc}(k_y\Delta y/2)\,\mathrm{sinc}(k_z\Delta z/2)\,]^{m+1}$
and $\boldsymbol{k}_{\boldsymbol{p}} = (k_x + 2\pi p_x/\Delta x,\; k_y + 2\pi p_y/\Delta y,\; k_z + 2\pi p_z/\Delta z)$, so that
$S_m^2(\boldsymbol{k}_{\boldsymbol{p}})S_m^2(\boldsymbol{k}_{\boldsymbol{p'}})/k_{\boldsymbol{p}}k_{\boldsymbol{p'}}\sim 1 /|\boldsymbol{p}|^{2m+3}\,|\boldsymbol{p'}|^{2m+3}$
at large $|\boldsymbol{p}|, |\boldsymbol{p'}|$. The term $\boldsymbol{p} = \boldsymbol{p'} = 0$ therefore dominates, with the remaining terms falling off, especially for higher spline order $m$.

\begin{figure}[H]
\includegraphics[width=0.98\columnwidth]{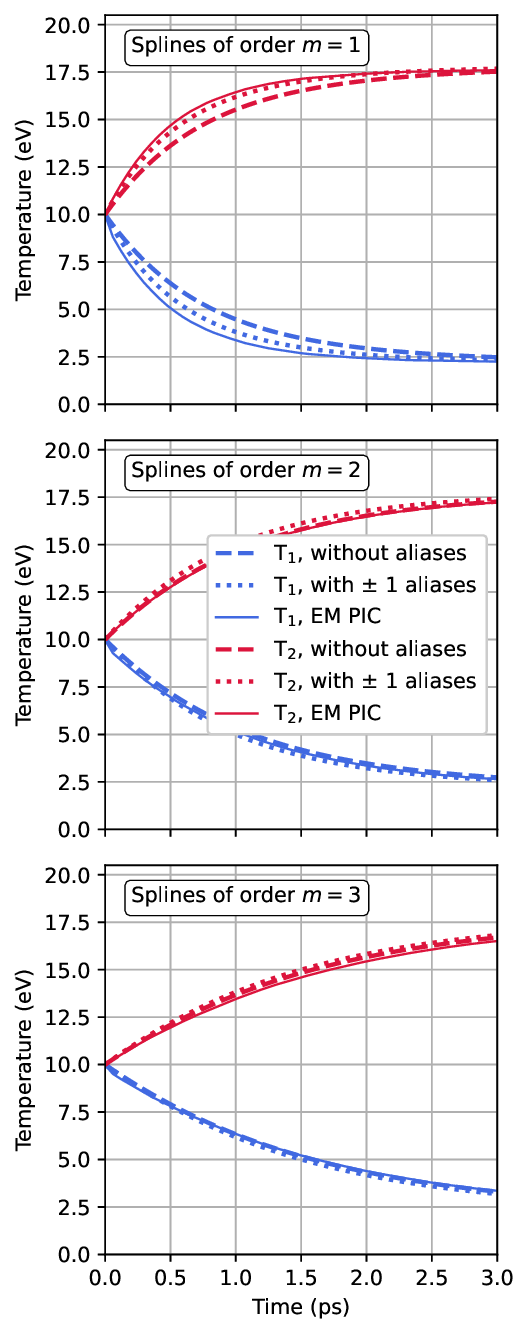}
\caption{Comparison between electromagnetic PIC simulations (EM PIC, solid lines) and the predictions of \cref{eq:evolutionT_anyD_with_aliases,eq:logPIC_anyD_with_aliases,eq:dielectricM_anyD_with_aliases}, using macroparticle shapes with different spline orders $m$ ($m=1,2,3$, from top to bottom). The dashed lines are obtained by retaining only the term $\boldsymbol{p} = \boldsymbol{p'} = \boldsymbol{p''} = 0$ in the sums over spatial aliases, while the dotted lines are obtained by retaining the first-order aliases, i.e. all terms with $p_x, p_y, p_z \in \{-1, 0, +1\}$ (and likewise for $\boldsymbol{p'}$ and $\boldsymbol{p''}$). The PIC simulations use the parameters of \cref{tab:parameters} and case B in \cref{tab:parameters3D}. The blue lines correspond to the electrons (species 1), and the red lines correspond to the ions (species 2). \label{fig:shape_factor_aliases} }
\end{figure}

To assess the importance of aliases, in \cref{fig:shape_factor_aliases} we evaluate \cref{eq:logPIC_anyD_with_aliases,eq:dielectricM_anyD_with_aliases} with the sums truncated so as to retain $p_x, p_y, p_z \in \{-1, 0, +1\}$, and likewise for $\boldsymbol{p'}$ and $\boldsymbol{p''}$ (dotted lines). These are compared with the predictions of the same equations when only the term $\boldsymbol{p} = \boldsymbol{p'} = \boldsymbol{p''} = 0$ is retained (dashed lines, identical to those in \cref{fig:shape_factor}), and with the PIC results (solid lines, also identical to those in \cref{fig:shape_factor}). For $m=1$ (top panel), the prediction including the first-order aliases lies much closer to the PIC results than the prediction without aliases, which confirms that the discrepancy observed for $m=1$ in \cref{fig:shape_factor} stems primarily from the neglect of spatial aliases. For $m=2$ and $m=3$, the dotted and dashed lines are nearly indistinguishable, consistent with the argument above that the contribution of aliases falls off rapidly with increasing spline order $m$.

\bibliographystyle{unsrt} 
\bibliography{sample}

\end{document}